\documentclass{aa}
\usepackage{longtable,booktabs}
\usepackage[varg]{txfonts}
\usepackage{latexsym}
\usepackage{amssymb}
\usepackage{lscape}
\usepackage[]{natbib}
\usepackage{subfigure}
\usepackage{xcolor}
\usepackage{multicol}
\usepackage{multirow}
\usepackage{amsmath}
\usepackage{booktabs}
\usepackage{caption}
\usepackage{pdflscape}
\usepackage{graphicx}
\usepackage{txfonts}
\def\lsim{\mathrel{\rlap{\lower 3pt \hbox{$\sim$}} \raise 2.0pt \hbox{$<$}}}
\def\gsim{\mathrel{\rlap{\lower 3pt \hbox{$\sim$}} \raise 2.0pt \hbox{$>$}}}

\begin{document}

\title{An extremely X--ray weak blazar at z=5}

\subtitle{}

\author{S.Belladitta\inst{1,2}                               
\and A. Moretti\inst{1}
\and A. Caccianiga\inst{1}                                                                  
\and G. Ghisellini\inst{3} 
\and C. Cicone\inst{1} 
\and T. Sbarrato\inst{4,3} 
\and L. Ighina\inst{1,4} 
\and M. Pedani\inst{5}
}

\institute{INAF - Osservatorio Astronomico di Brera, via Brera, 28, 20121 Milano, Italy\\
\email {silvia.belladitta@inaf.it}
\and
DiSAT, Universit\`a degli Studi dell'Insubria, Via Valleggio 11, 22100 Como, Italy
\and
INAF - Osservatorio Astronomico di Brera Merate, via E. Bianchi, 46, 23807 Merate, Italy
\and 
Dipartimento di Fisica G. Occhialini, Universit\`a degli Studi di Milano Bicocca, Piazza della Scienza 3, 20126 Milano, Italy
\and
INAF - Fundaci\'on Galileo Galilei, Rambla Jos\'e Ana Fernandez P\'erez 7, 38712 Bre\~{n}a Baja, TF, Spain
}

\date{Received; accepted}

\abstract
{We present the discovery and properties of DESJ014132.4-542749.9 (DES0141-54), a new powerful radio-loud active galactic nucleus (AGN) in the early Universe (z=5.0).
It was discovered by cross-matching the first data release of the Dark Energy Survey (DES DR1) with the Sidney University Molonglo Survey (SUMSS) radio catalog at 0.843 GHz.
This object is the first radio-loud AGN at high redshift discovered in the DES.
The radio properties of DES0141-54, namely its very large radio-loudness (R>10$^{4}$), the high radio luminosity (L$_{0.8 GHz}$=1.73$\times$10$^{28}$ W Hz$^{-1}$), and the flatness of the radio spectrum ($\alpha$=0.35) up to very high frequencies (120 GHz in the source's rest frame), classify this object as a blazar, meaning, a radio-loud AGN observed along the relativistic jet axis. 
However, the X--ray luminosity of DESJ0141-54 is much lower compared to those of the high redshift (z$\geq$4.5) blazars discovered so far. 
Moreover its X-ray-to-radio luminosity ratio (log($\frac{L_{[0.5-10]keV}}{L_{1.4GHz}}$)=9.96$\pm$0.30 Hz) is small also when compared to lower redshift blazars: only 2\% of the low-z population has a similar ratio. 
By modeling the spectral energy distribution we found that this peculiar X--ray weakness and the powerful radio emission could be related to a particularly high value of the magnetic field.
Finally, the mass of the central black hole is relatively small (M$_{BH}$ = 3-8 $\times$10$^8$ M$_{\odot}$) compared to other confirmed blazars at similar redshift, making DES0141-54 the radio-loud AGN that host the smallest supermassive black hole ever discovered at z$\geq$5. 
}

\keywords{Galaxies:  high-redshift  -- Galaxies: supermassive black holes -- Quasar: individual: DESJ014132.4-542749.9}
\maketitle

%

\section{Introduction}
Quasars (QSOs, or active galactic nuclei, AGNs) are powered by accretion of matter onto supermassive black holes (SMBHs, $\sim$10$^{6}$-10$^{10}$ M$_{\odot}$, e.g. Rees 1984).
They are the most luminous, non-transient sources in the sky, and hence they can be observed at extremely large cosmological look-back times (z>6, <1 Gyr after the Big Bang) across the entire electromagnetic spectrum.
Studying quasars at high redshift allows us to understand the formation and growth of SMBHs (e.g. Mortlock et al. 2011) and of their host galaxies (e.g. Fabian 2012), to learn about the chemical composition of the intergalactic medium (e.g. Jiang et al. 2016), and of the densest cosmic environments (e.g. Venemans et al. 2007), to probe the role of AGNs in driving the cosmic re-ionization process (e.g. Fan et al. 2006), and to test cosmological models (e.g. Cao et al. 2017).  
Nevertheless, the rarity of bright z>4.5 quasars and the high contamination by cool dwarf stars with similar colors, make high redshift AGNs difficult to find. 
In addition, wide-field surveys available today are affected by a limited sensitivity.\\
At low redshift, $\sim$10\% of AGNs are strong radio sources, meaning, radio-loud (RL) AGNs.
The origin of AGN radio emission is synchrotron radiation, which is produced by charged particles accelerated and collimated relativistically in a strong magnetic field, mostly along bipolar jets emitted from the central SMBH (e.g. Bridle \& Perley 1984; Zensus 1997).
The RL fraction does not appear to depend on redshift (e.g. Anderson et al. 2001, Ivezi{\'c} et al. 2002, Ba\~{n}ados et al. 2015), although some works found evidence of an evolution both in redshift and luminosity (e.g. Jiang et al. 2007).
Among RL AGNs, the most distant known at present day is at z=6.21 (Willott et al. 2010).
Due to the rarity of these objects, among the entire AGN population, it is difficult to build up statistically significant sample of RL AGNs, especially at high redshift.\\
Particularly important is the sub-class of blazars, meaning RL AGNs whose relativistic jet is seen at a small angle from our line of sight.
Blazars show a compact radio structure at milli-arcsecond (mas) scale and a high brightness temperature (Tb) larger than the inverse Compton equipartition limit (Tb>10$^{10}$ K, e.g. Readhead 1994, Homan et al. 2006) in their center.  
Since the radiation produced by relativistic jets is strongly boosted along the jet direction, blazars are very bright and are visible up to high redshifts, allowing the study of the RL population across cosmic time. 
Moreover, the peculiar orientation of blazars has a high statistical relevance: for each observed blazar there must be $\sim$2$\Gamma$$^{2}$ sources whose jets are pointing elsewhere (Volonteri et al. 2011, Ghisellini et al. 2013), where $\Gamma$ is the bulk Lorentz factor of the emitting plasma (typically, $\Gamma$ $\sim$10-15, e.g., Ghisellini et al. 2010). 
Therefore with blazars we can properly trace the entire population of RL AGNs.
To date, the most distant blazar known is at z=5.47 (Romani et al. 2004) and only a few others (i.e., five) have been discovered at z$\geq$5.0\footnote{The criteria adopted to classify a source as a blazar are not unique and they can be based on both radio and X--ray properties. Here the six sources considered as blazars at z$\geq$5.0 are: J0906+6930 (z=5.47, Romani et al. 2004), J1648+4603 (z=5.38, Caccianiga et al. 2019), J1026+2542 (z=5.25, Sbarrato et al. 2012),  J0131-0321 (z=5.18, Ghisellini et al. 2015), J1146+4037 (z=5.005, Ghisellini et al. 2014), and J1629+1000 (z=5.00, Caccianiga et al. 2019). Among these, however, only J0906+6930, J1026+2542, and J0131-0321 are confirmed blazars also from evidence of Doppler boosting with Very Long Baseline Interferometry, VLBI, observations.}. \\ 
Studying the properties of these high-z sources and comparing them to those of the low redshift blazar population is important to understanding their evolution across the cosmic time.
In an effort to identify and study the high-z RL AGN population, we are conducting a project that combines optical, infrared (IR), and radio datasets to identify radio sources at z>4.0 (see Caccianiga et al. 2019 for the first results of this project, using the CLASS sample, in the northern hemisphere).
In this paper we describe the discovery of DESJ014132.4-542749.9, a new blazar candidate at z=5.0 selected from the combination of the optical Dark Energy Survey (DES, Flaugher et al. 2005) and the radio Sydney University Molonglo Sky Survey (SUMSS, Bock et al. 1999, Mauch et al. 2003). \\
The paper is structured as follows: in Section \ref{selection} we describe the selection method based on archival data; in Section \ref{obsred} we show our new dedicated observations of 
DESJ014132.4-542749.9 in the optical and X--ray bands; the results of our work and the discussion are presented in Sections \ref{results}--\ref{sedmodel}; finally, in Section \ref{sumconc}, we summarize our conclusions.\\
The magnitudes used in this work are all in the AB system. 
We use a flat Lambda cold dark matter ($\Lambda$CDM) cosmology with H$_0$=70 km s$^{-1}$ Mpc$^{-1}$, $\Omega_m$=0.3 and $\Omega_\Lambda$=0.7. 
Spectral indexes are given assuming S$_{\nu} \propto$ $\nu^{-\alpha_{\nu}}$ and all errors are reported at 1$\sigma$, unless otherwise specified.

\section{Candidate selection}
\label{selection}
\begin{figure}[!h]
        \centering
        \includegraphics[width=8.5cm]{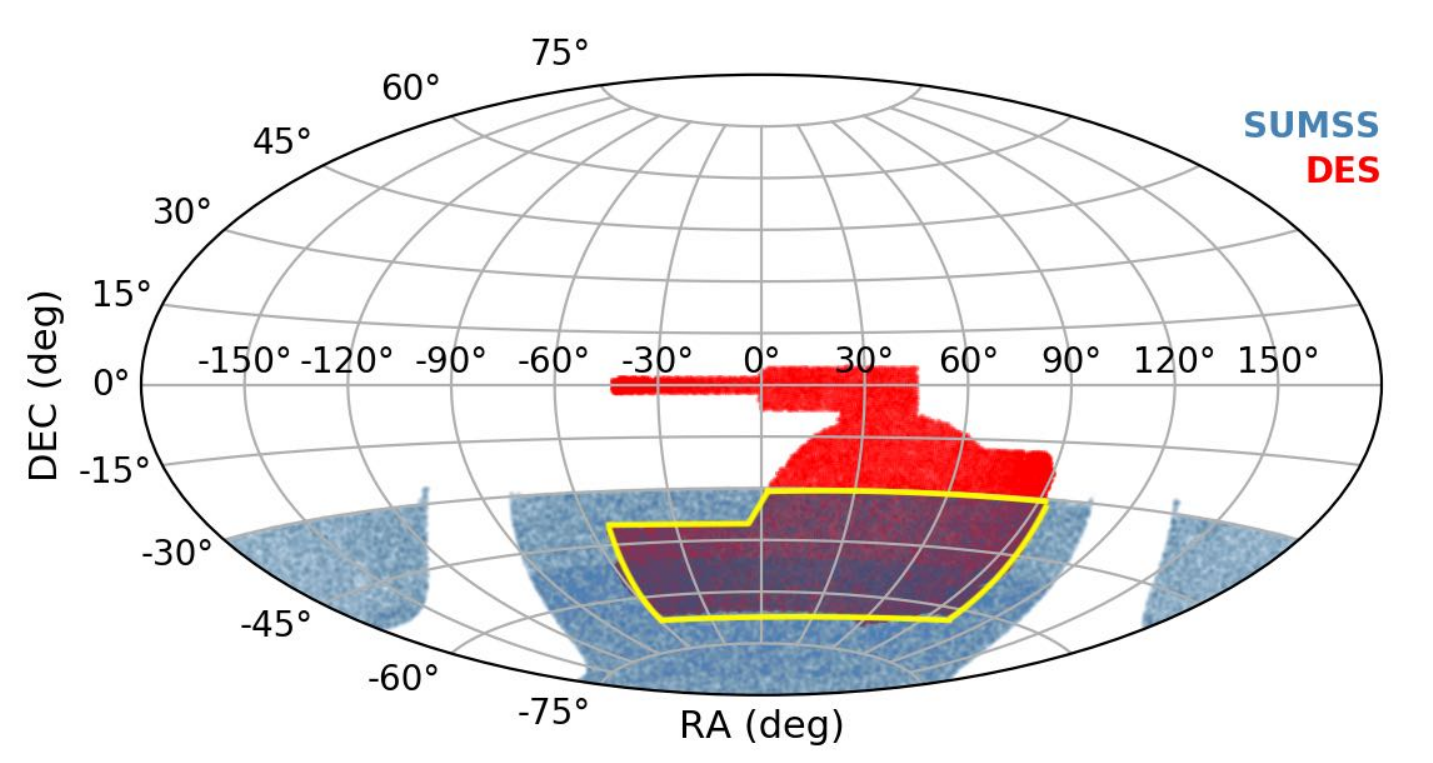}
        \caption{DES (in red) and SUMSS (in light blue) sky coverage. The common area of the two surveys in which we searched for high-z RL AGNs is 3500 deg$^2$ (yellow box).}
        \label{skymap}
\end{figure}

\begin{figure*}[!h]
        \centering
        {\includegraphics[width=3.0cm]{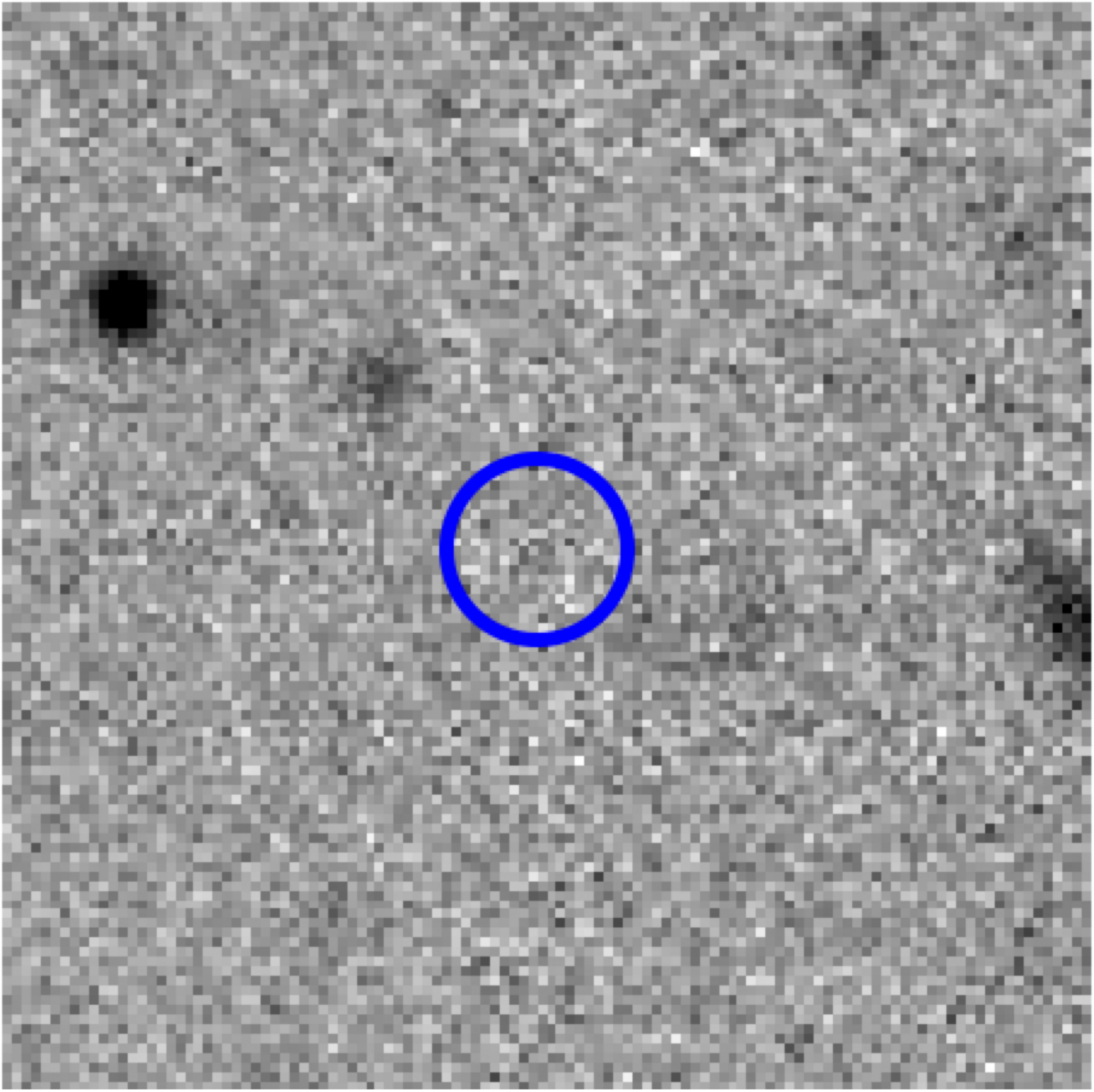}\hspace{-0.3mm}
                \includegraphics[width=3.0cm]{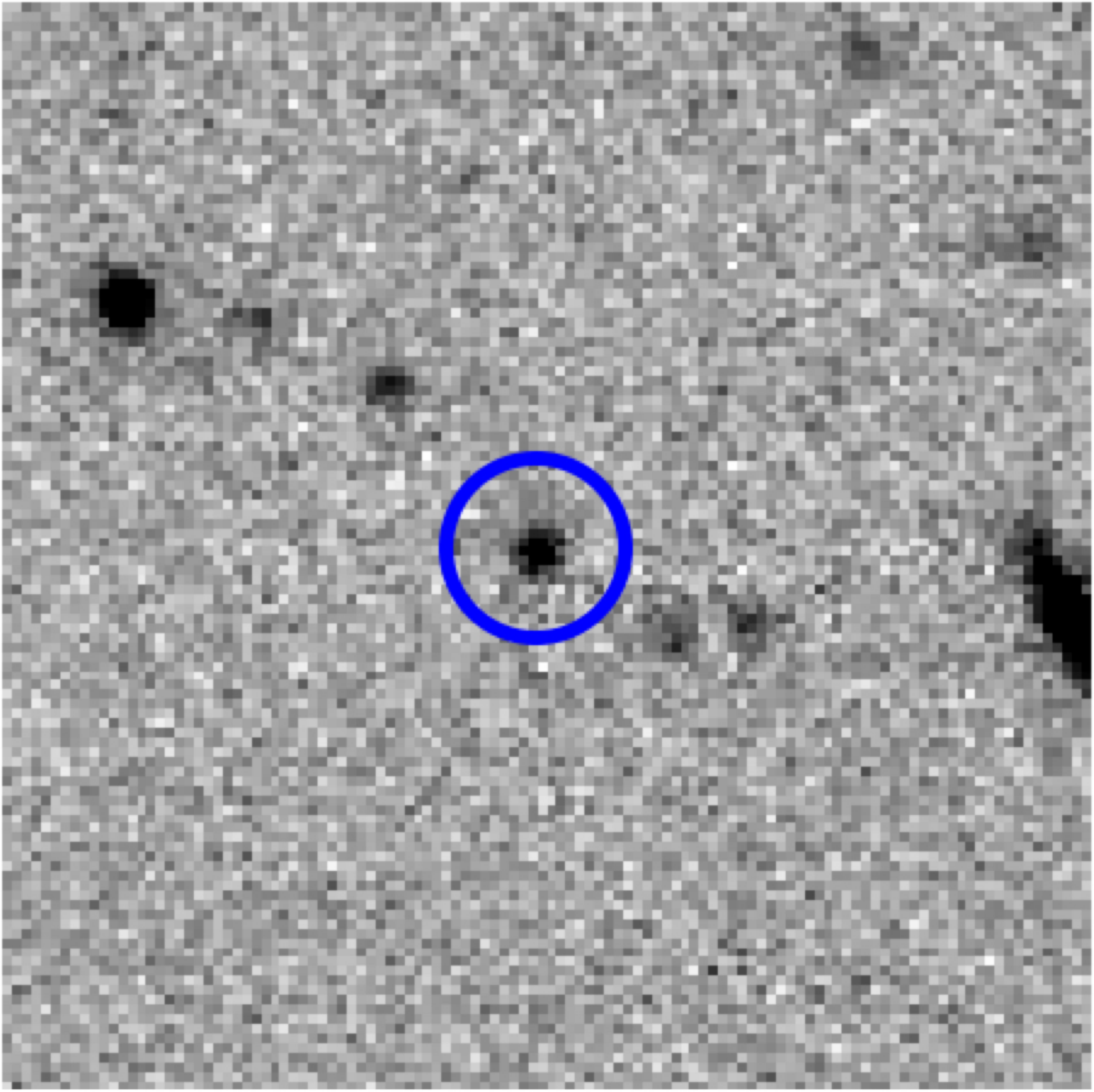}\hspace{-0.3mm}
                \includegraphics[width=3.0cm]{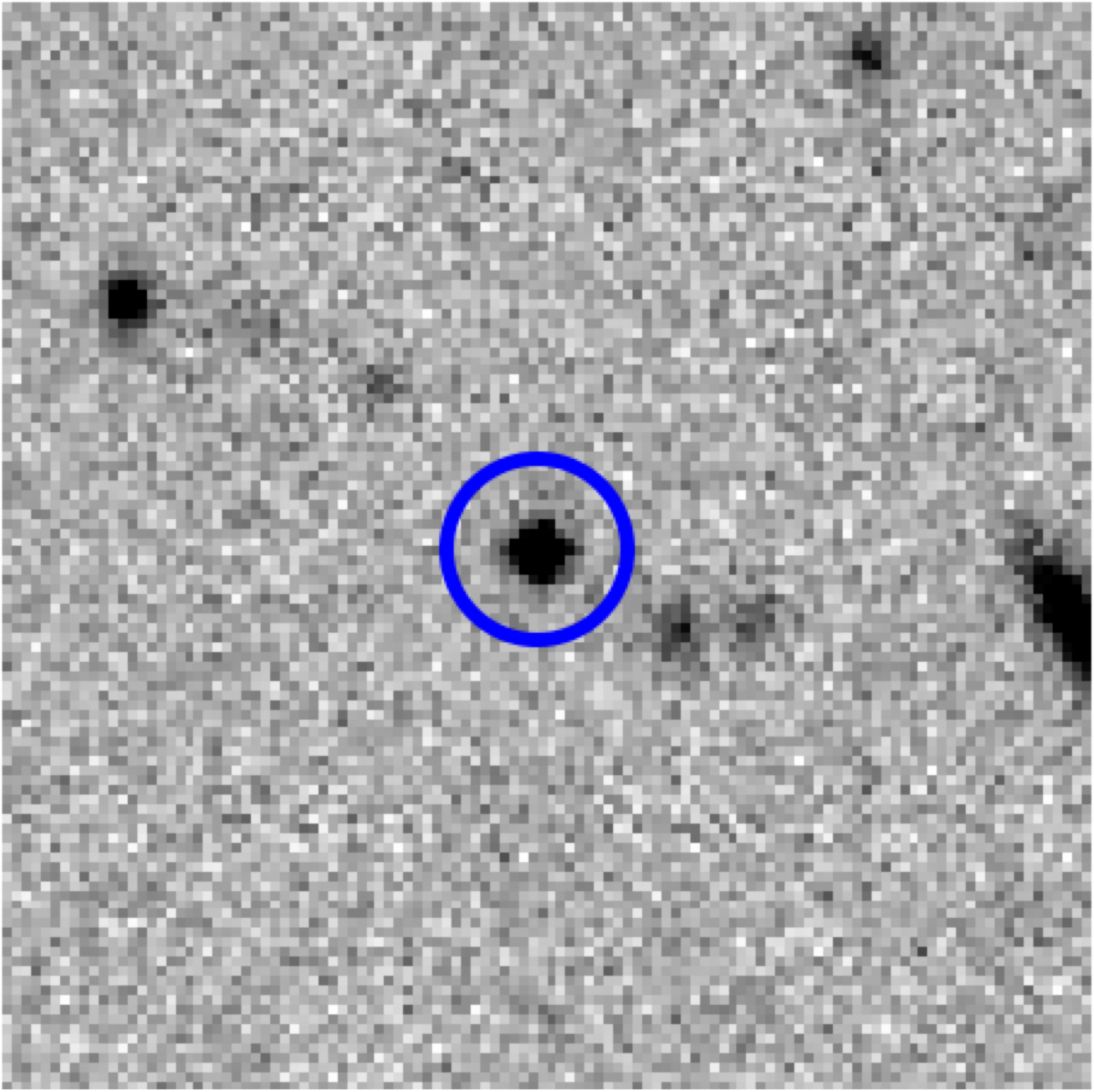}\hspace{-0.3mm}
                \includegraphics[width=3.0cm]{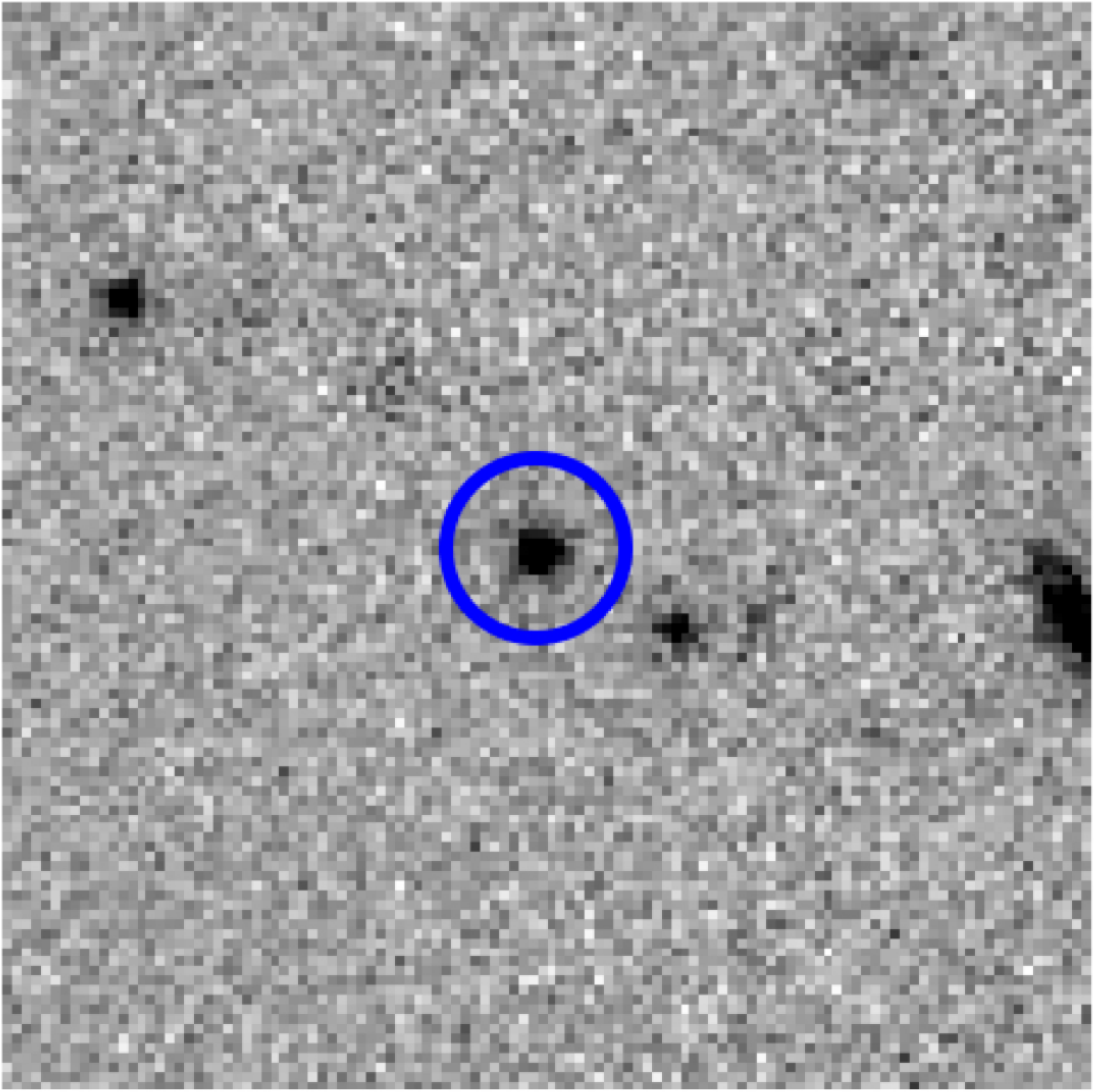}\hspace{-0.3mm}
                \includegraphics[width=3.0cm]{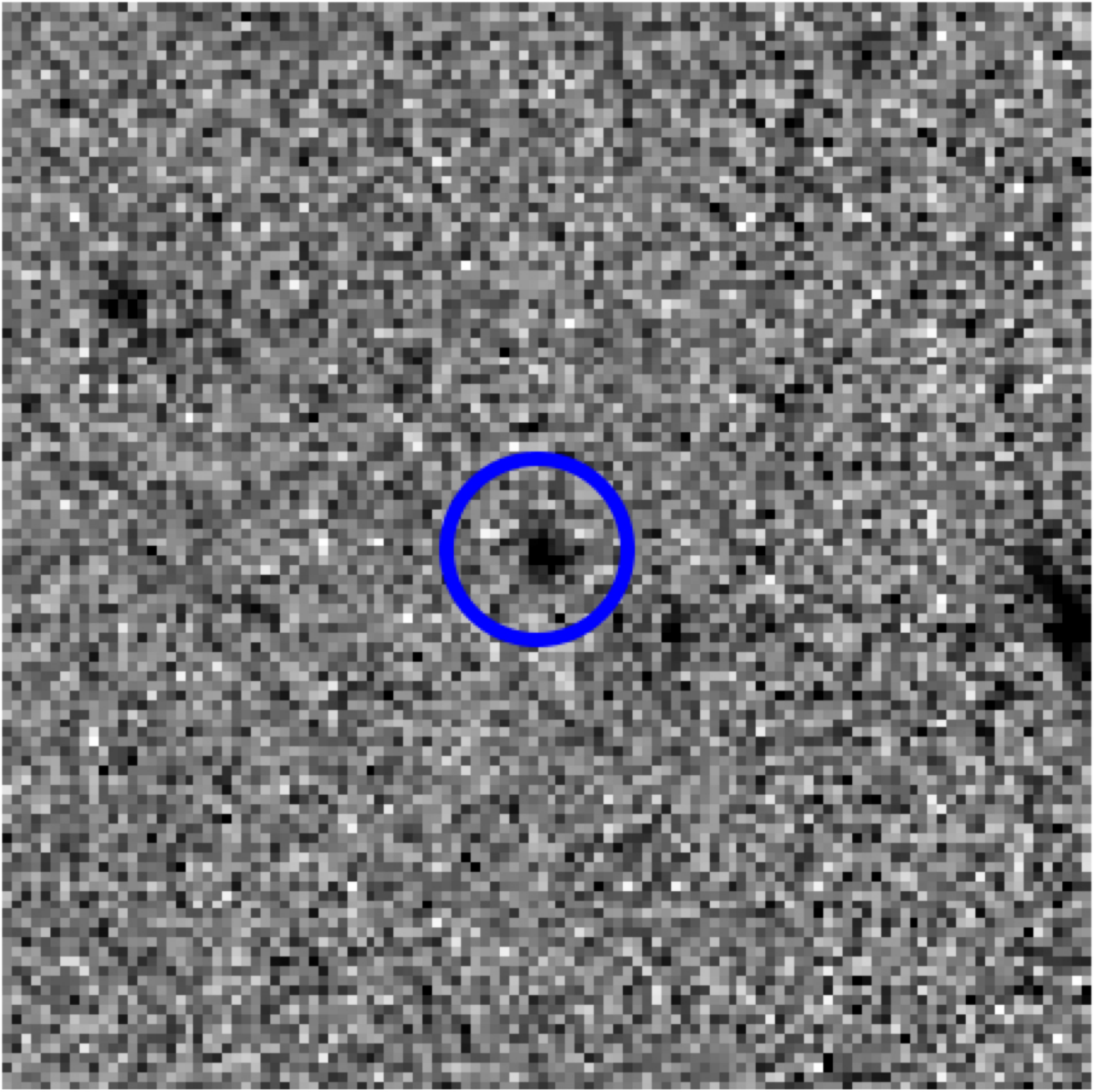}\hspace{-0.3mm}\
            \includegraphics[width=3.0cm]{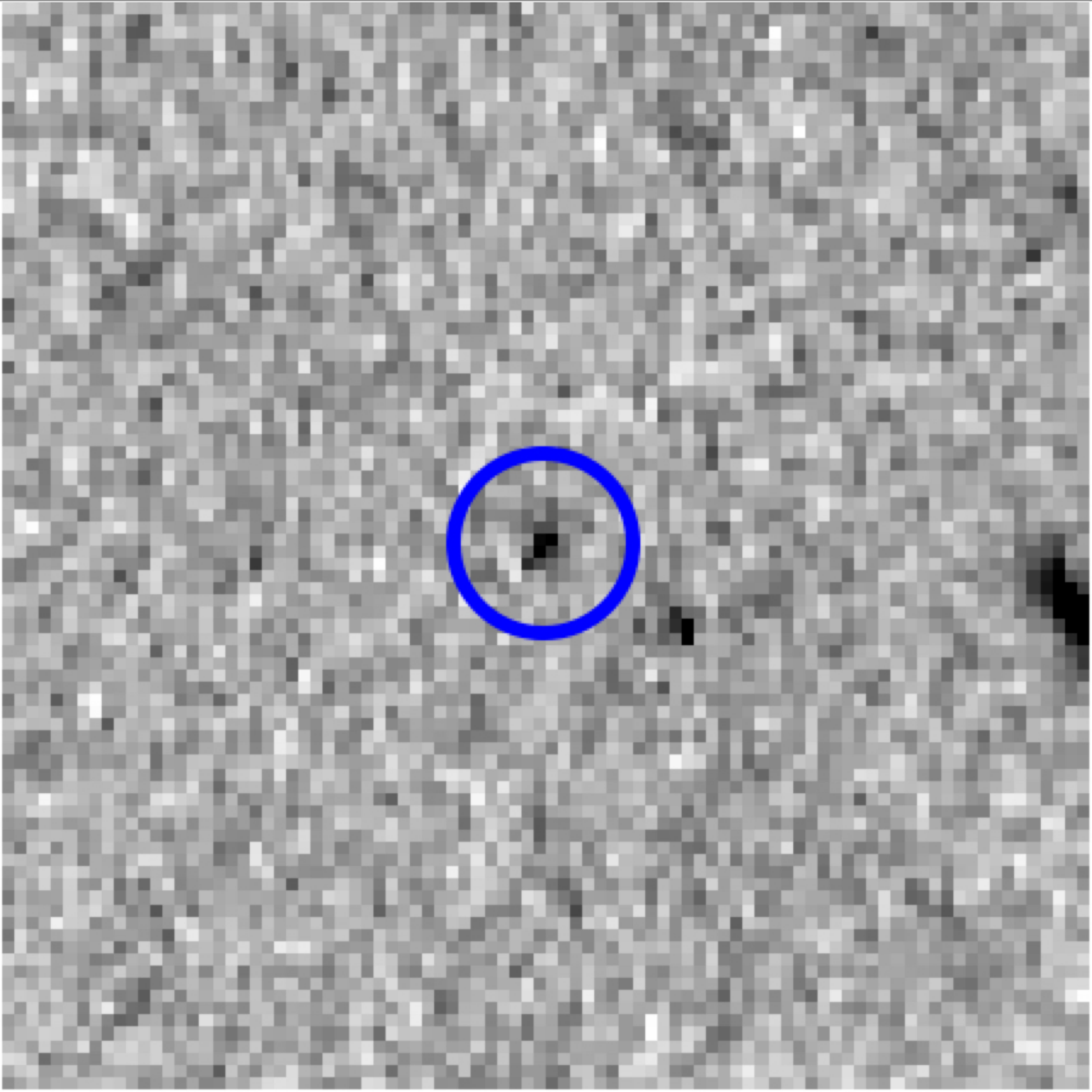}}
        \vspace{-0.2cm}
        \caption{0.5$\arcmin$$\times$0.5$\arcmin$ DES $g,r,i,z,Y,$ and VHS $J$ cutout images of DES0141-54. The object position is marked in all the images with a blue circle of 2.5$\arcsec$ in diameter. All images are oriented with north to the top and east to the left.}
        \label{photimag}
\end{figure*}

To search for high-z blazars in the southern hemisphere we exploited the combination of the DES first data release (DR1, Abbot et al. 2018) and the SUMSS radio catalog.
DES covers 5000 deg$^2$ of the southern Galactic cap in five broad photometric bands, $grizY$, using the Dark Energy Camera (DECam; Honscheid et al. 2008, Flaugher et al. 2015) mounted on the Blanco 4-meter telescope at the Cerro Tololo Inter-American Observatory (Chile). 
The nominal median co-added catalog depths at signal-to-noise ratio (S/N) = 10 are $g$ = 24.33, $r$ = 24.08, $i$ = 23.44, $z$ = 22.69, and $Y$ = 21.44 mag (Abbott et al. 2018). 
DES is about two orders of magnitude deeper than the Sloan Digital Sky Survey (SDSS, York et al. 2000) and the Panoramic Survey Telescope and Rapid Response System (Pan-STARRS1, Chambers et al. 2016) (r$\sim$22.8). 
Therefore this survey offers new possibilities to discover high-z QSOs (both RL and radio-quiet, e.g. Reed et al. 2017, Reed et al. 2019) that are less luminous, and therefore, on average, with a smaller SMBH, with respect to those found with previous shallower surveys.
DES is partially covered by SUMSS, a radio imaging flux limited (S$_{lim}$ = 2.5 mJy) survey of the southern sky at $\delta$ < -30$^{\circ}$, covering a total area of 8100 deg$^2$, carried out at 843 MHz with the Molonglo Observatory Synthesis Telescope (MOST, Mills 1981, Robertson 1991) using a spatial resolution (beam) of 43$\arcsec$.
The intersection between these two surveys covers 3500 deg$^2$ of the sky (see Fig. \ref{skymap}).\\
To search for high-z RL candidates in this area we used the color selection technique (i.e., the dropout method, e.g., McGreer et al. 2013, Mazzucchelli et al. 2017, Caccianiga et al. 2019).
In particular we used the $r-i$ versus $i-z$ color diagram to select candidates at z$\sim$5.
For our selection we considered a list of 91301 sources from the DES DR1 database with these characteristics:
\begin{itemize}
\item inclusion in the SUMSS catalog: $\delta$ < -31$^{\circ}$ 
\item observability with ESO New Technology Telescope: mag$_{auto\_i}$<22.5
\item color selection: mag$_{auto\_r}$ - mag$_{auto\_i}$>1.5 and mag$_{auto\_i}$ - mag$_{auto\_z}$<0.5
\item no detection in $g$ band: mag$_{auto\_g}$>25.0 and mag$_{aper2\_g}$ > 25.0
\item reliable magnitudes: magerr$_{auto\_z}$<0.2, magerr$_{auto\_i}$<0.2 and magerr$_{auto\_y}$<0.3\footnote{This last constraint has been placed to have more reliable colors.}
\end{itemize}
where mag$_{auto}$ is the magnitude estimate for an elliptical model based on the Kron radius and mag$_{aper2}$ is the magnitude estimate for circular apertures of 2$\arcsec$.\\
We then cross-correlated this list with the SUMSS catalog of very bright (S$_{0.8 GHz}$ > 30 mJy) and relatively compact (S$_{INT}$/S$_{Peak}$ < 1.5) objects (71254 objects), which have higher chances to have their relativistic jet pointing toward Earth.
We used a maximum separation equal to the SUMSS radio positional error (err = $\sqrt{(err_{RA}^{2}+err_{DEC}^{2})}$) for each source. 
Positions in the SUMSS catalog are accurate to within 1-2$\arcsec$ for objects with peak brightness $\geq$20 mJy/beam (at 90\% level of confidence), and are always better than 5$\arcsec$ (Bock et al. 1999).\\
With this selection we found three candidates, but only  DESJ014132.4-542749.9 (hereafter DES0141-54) has a spectral energy distribution (SED) consistent with that of a non-absorbed quasar at high redshift. 
Therefore it was considered the only promising candidate for spectroscopic follow-up observations.
Its optical coordinates are: right ascension (RA) = 25.38517 deg, declination (Dec) = -54.46386 deg. 
Figure \ref{photimag} shows the optical images of DES0141-54 in the different $grizY$ bands.
The distance between the optical and SUMSS radio position of DES0141-54 is 2.2$\arcsec$.
At radio wavelengths, besides the SUMSS catalog, several radio detections are reported in other surveys (see Table \ref{Tradio} for details). 
In particular, DES0141-54 has been detected by the Australia Telescope 20 GHz Survey high-resolution follow up (AT20G$_{follow-up}$, Chhetri et al. 2013) and the Australia Telescope-PMN survey (ATPMN, McConnell et al. 2012), both carried out with the Austalia Telescope Compact Array (ATCA) using the longest baseline ($\sim$6km) possible, allowing a high angular resolution ($\sim$1$\arcsec$ at 8.4 GHz). 
From these detections we can infer that the radio-optical association of our source is reliable (the distance between the ATCA and the DES position is $\sim$0.15$\arcsec$).
 
\begin{footnotesize}
        \begin{table*}[!h]
                \caption{Radio detections of DES0141-54 found in radio archives.} 
                \label{Tradio}
                \centering
                \begin{tabular}{cccc}
                        \hline\hline
                        $\nu_{obs}$  & S$_{\nu}$ & reference survey & years of observation \\ 
                        (GHz) & (mJy) &  &  \\  
                        (1) & (2) & (3)  & (4) \\ 
                        \hline 
                        20       & 43.0$\pm$2.0    & AT20G  & 2004-2008 \\
                        
                        8.6      & 70.0$\pm$4.0    & AT20G$_{follow-up}$  & 2004-2008 \\
                        
                        8.6      & 63.0$\pm$10.0   & ATPMN  & 1992-1994 \\
                        
                        8.4      & 59.5$\pm$3.0   & CRATES & 2005 \\
                        
                        4.85     &100.0$\pm$9.0    & PMN & 1990 \\        
                        
                        4.8      & 90.0$\pm$5.0    & AT20G$_{follow-up}$ & 2004-2008 \\
                        
                        4.8      &  82$\pm$7.0     & ATPMN & 1992-1994 \\    
                        
                        0.843    &  178.3$\pm$5.5  & SUMSS & 2002 \\      
                        
                        0.076-0.231      & 392.5$\pm$69.8 - 241.9$\pm$12.3 & GLEAM & 2013-2014 \\               
                        \hline
                        
                \end{tabular}
                \tablefoot{Col(1): observed frequency in GHz; Col(2): flux density in mJy; Col(3) reference survey: AT20G = Australia Telescope 20 GHz Survey (Murphy et al. 2010), AT20G$_{follow-up}$ = the AT20G high-resolution follow-up (Chhetri et al. 2013), ATPMN = Australia Telescope-PMN survey (McConnell et al. 2012), CRATES = Combined Radio All-sky Targeted Eight GHz Survey (Healey et al. 2007), PMN = 4.85 GHz PARKES-MIT-NRAO survey (Wright et al. 1994),  SUMSS = Sidney University Molonglo Survey (Mauch et al. 2003), GLEAM = GaLactic and Extragalactic All-sky MWA Survey (Wayth et al. 2015, Hurley-Walker et al. 2017); Col(4) years of the observation.\\ In the last line we report the range of the GLEAM detections, from 76 to 231 MHz. All the GLEAM fluxes can be found in Table \ref{Tradio2}.}
        \end{table*}
\end{footnotesize}

\subsection{Archive photometry: infrared and X-rays data}
\label{wisexmm}
\begin{figure}[!h]
	\centering
	\includegraphics[width=3.5cm]{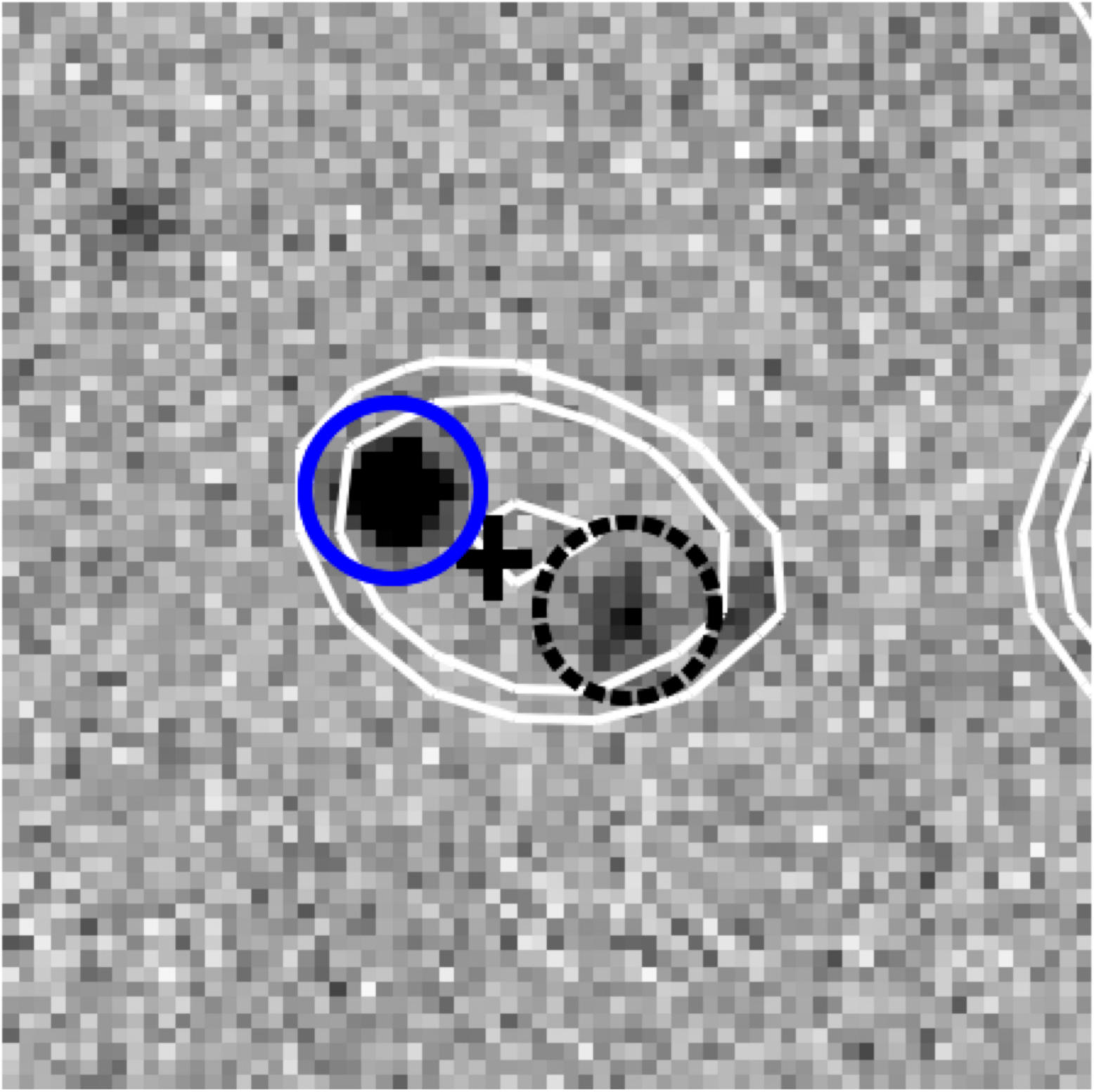}
	        \vskip -0.1cm
	\caption{WISE w1 contours (in white) overlapped on DES $i$--band 0.3$\arcmin$$\times$0.3$\arcmin$ image. The object position is marked with the same blue circle as from Fig. \ref{photimag}. The central cross marks the position of the WISE detection. The contours illustrate that WISE does not spatially resolve the emission of DES0141-54 and of a southwest source clearly visible in the optical bands (dashed black circle). 
		The image is oriented with north to the top  and east to the left.}
	\label{irmag}
\end{figure}

DES0141-54 is included in the AllWISE Source Catalog (WISE, Wright et al. 2010; NEOWISE, Mainzer et al. 2011), with clear detections (signal-to-noise ratio>4) at $\lambda$ = 3.4$\mu$m (w1) and $\lambda$ = 4.6$\mu$m (w2).
The WISE catalog magnitudes have been converted from Vega to the AB system. 
The offset between the DES optical and WISE IR positions is 2.06$\arcsec$.
However, the WISE detection is the result of blending of two distinct sources that are clearly visible in the optical images (see Fig. \ref{irmag}).
Therefore we corrected the WISE magnitudes used in this work for this effect with the following procedure. 
We modeled the WISE emission as the sum of two point-like sources at the optical positions of DES0141-54 and of the other object to the southwest, leaving the relative normalizations free to vary.
For a PSF reference model we used a point-like IR source nearby in the field of view. 
We estimated that the DES0141-54 contribution to the WISE flux is 60\% both in w1 and w2, and we corrected the magnitudes accordingly.\\
Moreover, a Vista Hemisphere Survey (VHS, McMahon et al. 2013) $J$ band detection is present in the VISTA Science Archive (VSA, Cross et al. 2012) at a distance of 0.073$\arcsec$ from the optical coordinates (last image on the right in Fig. \ref{photimag}). 
The optical and IR magnitudes of DES0141-54 are listed in Table \ref{Tmag}. \\
\newline
DES0141-54 was serendipitously observed by XMM-Newton in 2005, in the field of Abell2933 (OBsID:0305060101, revolution 1089), with a 8.7arcmin of offaxis. 
The effective exposure times are 31ks and 23ks for MOS and PN, respectively. 
The source is faint and it is not included in the 3XMM-DR8 catalog (Rosen et al. 2019).
We performed the data reduction using the interactive data analysis task of the XMM-Newton Science Archive (XSA), considering a circular region of 5$\arcsec$ radius centered on the optical position of DES0141-54.
The source is detected only in the PN image with eleven counts and a significance of $\sim$3$\sigma$.
We selected a small region to perform the analysis because of the presence of a nearby bright source (at $\sim$15$\arcsec$ to the northeast) and of the low counts detected for our object. 
Flux losses are at the level of 50\% and are accounted for the ancillary response file  (ARF) calculation.
Using XSPEC version 12.10.1 (Arnaud 1996), we modeled the PN data with a simple absorbed power law with the absorption factor fixed to the Galactic value (2.5$\times$10$^{20}$ cm$^{-2}$) as measured by the HI Galaxy map (Kalberla et al. 2005).
The spectrum was grouped with a minimum of one count for each bin, and the best fit was calculated using the C-statistics.
We measured a flux of 7.5$\pm$2.6$\times$10$^{-15}$ erg s$^{-1}$ cm$^{-2}$ for the [0.5-10]keV energy band, using the best fit value for the photon index ($\Gamma$=1.75$\pm$0.5).

\begin{footnotesize}
	\begin{table}[!h]
		\caption{Optical and near IR AB magnitude of DES0141-54.}
		\label{Tmag}
		\centering
		\begin{tabular}{cccc}
			\hline\hline
			filter & $\lambda$  & magnitudes & reference survey \\
			& ($\mu$m) & &  \\
			(1) & (2) & (3) & (4) \\
			\hline
			$r$       & 0.659 & 22.96$\pm$0.17 & DES \\
			
			$i$       & 0.789 & 21.16$\pm$0.12 & DES \\     
			
			$z$       & 0.976 & 21.18$\pm$0.15 & DES \\     
			
			$Y$       & 1.003  & 21.74$\pm$0.30 & DES \\
			
			$J$       & 1.254 & 21.14$\pm$0.28 & VHS \\     
			
			$w1$      & 3.4 & 20.24$\pm$0.08 & WISE \\      
			
			$w2$      & 4.6 & 20.80$\pm$0.29 & WISE \\      
			\hline
		\end{tabular}
		
		\tablefoot{Col(1) optical and near IR filters; Col(2) central wavelength in micron; Col(3) observed magnitudes: the $J$, w1, and w2 WISE magnitudes have been converted from Vega to the AB system. The WISE magnitudes have been also corrected for blending of a nearby source as described in Section \ref{wisexmm}; Col(4) reference survey.}
	\end{table}
\end{footnotesize}

\section{New observations and data reduction}
\label{obsred}

\subsection{NTT observations}
\label{NTT}
After its identification in the DES catalog (see Section \ref{selection}) we performed a dedicated spectroscopic follow-up of DES0141-54 with the ESO Faint Object Spectrograph and Camera (EFOSC2, Buzzoni et al. 1984) mounted at the New Technology Telescope (NTT), located at ESO-La Silla observatory, on 2018 February 9 during the 0100.A-0606(A) observing program (PI: A. Moretti). 
We carried out four 15–minute observations, to remove cosmic ray hits, with a long-slit of 1.5$\arcsec$ width, for a total exposure time of 1h.
The average seeing during the observations was 1.1$\arcsec$ and the average air mass was 1.5.
The wavelength calibration was secured by means of frequent exposures of a He-Ar hollow-cathode lamp.
The spectrograph response (i.e., the sensitivity function) was obtained by many exposures of LTT1788 (RA = 03:48:22.17, Dec = -39:08:33.6) spectrophotometric star from the catalog of Hamuy et al. (1992, 1994).
The EFOSC2 data reduction was performed using standard Image Reduction and Analysis Facility (IRAF) procedures (Tody 1993).  
Panel (a) of Fig. \ref{spec} shows the EFOSC2/NTT spectrum of DES0141-54, with the most important emission lines marked.
The object's redshift has been measured by fitting the Ly-$\alpha$$\lambda$1216Å line with a single Gaussian profile using the IRAF task $splot$.
The resulting value is z=5.00$\pm$0.01, confirming the high redshift nature of the selected candidate. 

\subsection{X-Shooter follow-up}
\label{Xsho}
\begin{figure*}
        \centering
        \subfigure[]
        {\includegraphics[width=9.5cm,height=7.5cm]{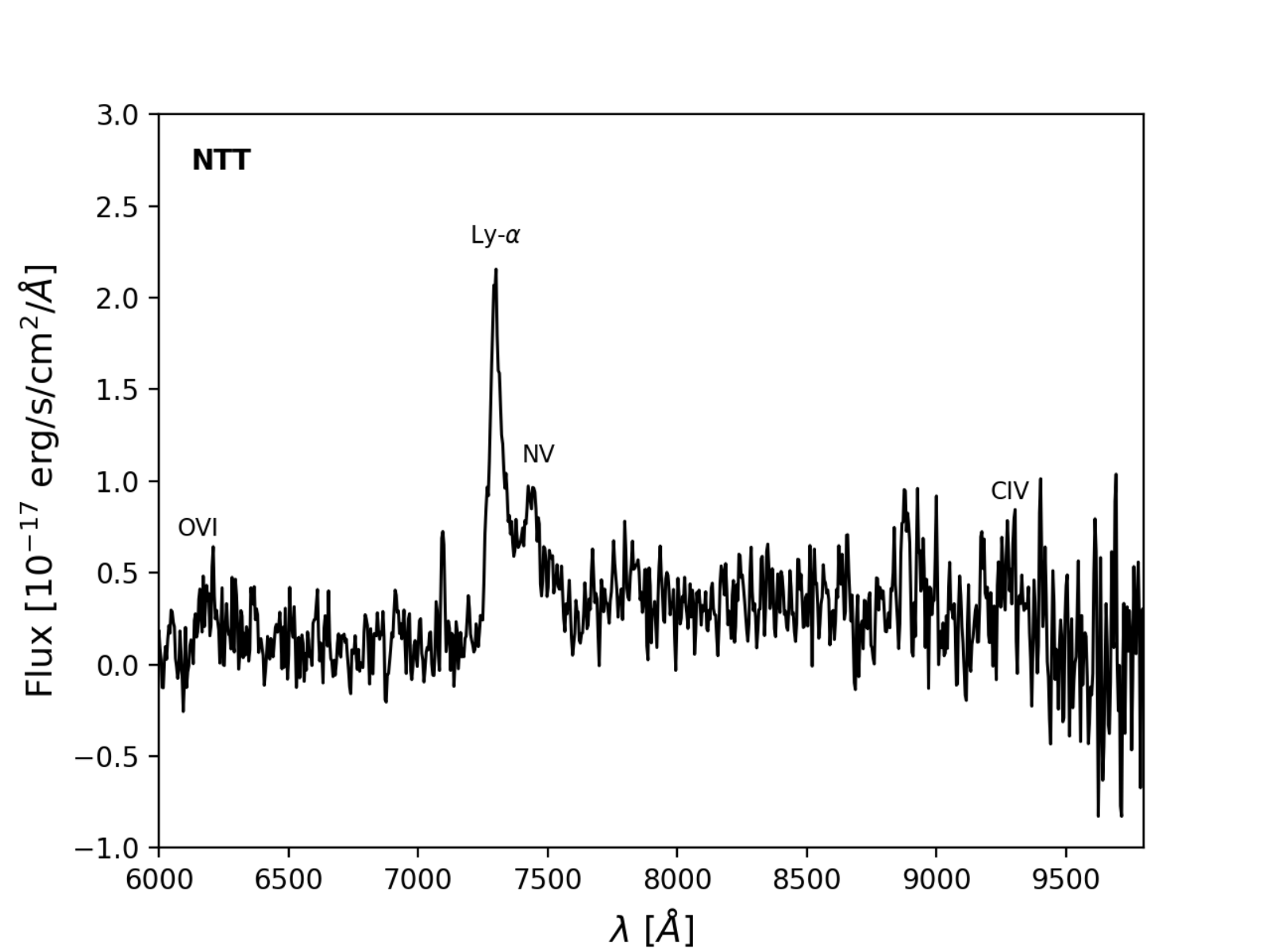}}
        \hspace{-9.0mm}
        \subfigure[]
        {\includegraphics[width=9.5cm,height=7.5cm]{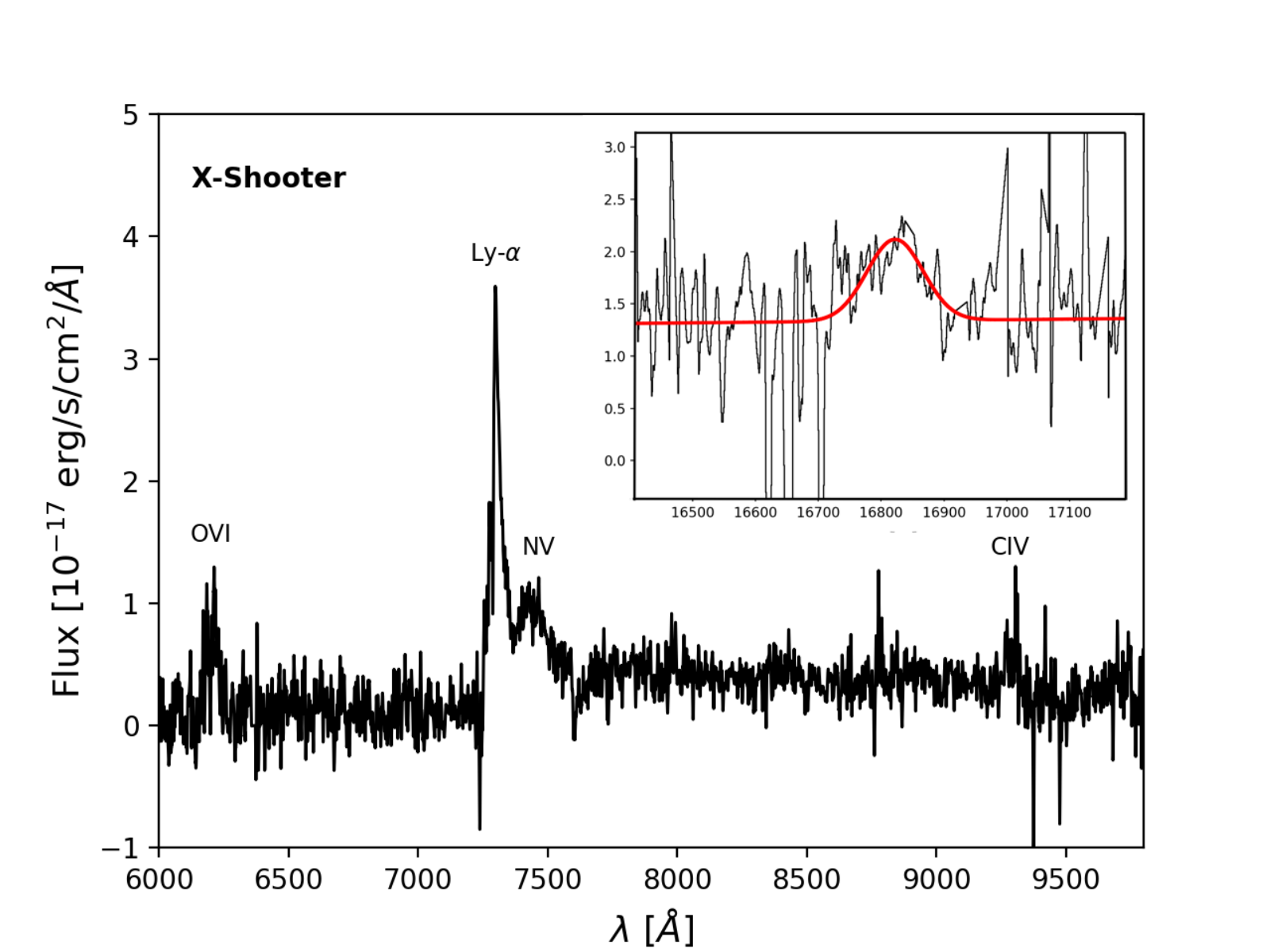}}
        \caption{Optical spectra of DES0141-54. Left panel: EFOSC2/NTT discovery spectrum; Right panel: X-Shooter/VLT follow-up. In the small inset on the right the zoomed-in portion of the IR spectrum with the Gaussian fit of the Mg$\rm{II}$$\lambda$2798Å line. In both panels the Ly-$\alpha$$\lambda$1216Å, the N$\rm{V}$$\lambda$1242.80Å, the C$\rm{IV}$$\lambda$1549Å and the O[VI]$\lambda$1035Å emission lines are shown.}
        \label{spec}
\end{figure*}

We performed a spectroscopic follow-up with X-Shooter at the ESO Very Large Telescope (VLT) in order to extend the wavelength range in the near-IR band and detect the Mg$\rm{II}$$\lambda$2798Å emission line, which can be used to estimate the central black hole mass of DES0141-54.
The observation was carried out in a Director's Discretionary Time program (project ID: 2100.A-5039, PI: A. Moretti) on 2018 February 23 and consisted of four exposures of 15 minutes each in nodding mode in the sequence ABBA, with a total integration time of 1h, simultaneously in the ultraviolet (UVB, 3100–5500 Å), visible (VIS, 5500–10150 Å) and near-IR (NIR, 10150–24800 Å) spectroscopic arms.  
The observation was performed along the parallactic angle on a clear night with an average seeing of 0.8$\arcsec$ and a mean air mass of 1.8.
The adopted slit widths were 1.0$\arcsec$ in the UVB, and 0.9$\arcsec$ in the VIS and NIR.  
The object was observed using a $K$-band blocking filter that lowers the sky background where scattered light from the $K$-band affects the $J$-band (Vernet et al. 2011). 
The spectrophotometric GD71 star (R.A = 05:52:27.51, Dec = +15:53:16.6) from the catalog of Moehler et al. (2014) was observed on the same night to ensure the flux calibration, and a series of exposures of a ThAr hollow-cathode lamp were done for wavelength calibration.
The data were reduced with the ESO/X-Shooter pipeline v2.5.2 (Modigliani et al. 2010) using the EsoRex interface\footnote{http://www.eso.org/cpl/esorex.html.} (Goldoni et al. 2006). 
However, we noticed that the reduced NIR spectrum shows systematically large and frequent sky-subtraction residuals, due to the IR Earth background that is difficult to remove, because it changes in space and time. 
Consequently, we opted to use some IRAF tasks to improve the reduction: $background$ for the sky subtraction and the tasks for the flux calibration ($standard$, $sensfunc,$ and $calibrate$). 
The optical X-Shooter spectrum is shown in panel (b) of Fig. \ref{spec}, in which we also report in the small inset, the IR spectrum including the Mg$\rm{II}$$\lambda$2798Å line with the Gaussian fit overlapped.
Despite the several residual lines in the IR spectrum, we detected the Mg$\rm{II}$$\lambda$2798Å line at a S/N$\sim$2.
The redshift was measured, as it was for the EFOSC2/NTT spectrum, by fitting a Gaussian profile on the Ly-$\alpha$ line using the IRAF task $splot$: the extracted value is z=5.005$\pm$0.003, confirming the previous measured value. 
The redshift was also confirmed by fitting the other detected lines with Gaussian profiles: O[VI]$\lambda$1035Å, N$\rm{V}$$\lambda$1242.80Å and the C$\rm{IV}$$\lambda$1549Å. We obtained z=4.998$\pm$0.004, z=4.999$\pm$0.002 and z=4.999$\pm$0.003 respectively.
For our analysis we used as the best estimator the redshift computed from the average of these values: z=5.000$\pm$0.002.

\subsection{X-ray follow-up with $Swift$}
\label{xray2}
The X--ray flux of DES0141-54 extracted from the XMM-Newton image is faint compared to what we expect from a blazar source at high redshift (see Section \ref{xrayprop} for details.). 
A possibility is that the X--ray flux of the source has varied. 
Indeed, blazars are known to be highly variable sources at all wavelengths, including the X--rays (e.g. Giommi et al. 2019).
Therefore, in order to check for possible variability of the X-ray emission of DES0141-54 we asked for a $Swift$-XRT (Gehrels et al. 2004) pointed observation.
The observation was carried out with a Target of Opportunity (ToO) request (target ID: 10586, PI: A. Moretti) on 2018 March 2 with a total exposure time of 8340s.
The integration would have been long enough to detect the source if 
it had an X--ray flux similar to that of the other blazars at z$\geq$5.0 reported in the literature (e.g. Sbarrato et al. 2012).
However, this observation has resulted in a non detection. 
We computed an upper limit of 1.3$\times$10$^{-14}$ erg s$^{-1}$ cm$^{-2}$ in the [0.5-10]keV energy band, assuming an N$_{H}$ and a $\Gamma$ equal to those used for the XMM-Newton analysis. 
This upper limit is consistent with the faint flux computed from the XMM-Newton image, confirming the weakness of DES0141-54 X--ray emission.

\section{Results}
\label{results}
In this section we report the results of our analysis on the multiwavelength properties of DES0141-54, by comparing them with those of other high-z RL AGNs.\footnote{For clarity: we define "radio-loud" (or RL) an AGN with a radio-loudness, as defined in Shen et al. (2011), larger or equal than 10.} 
From the literature we collected radio fluxes, optical-IR magnitudes and X--ray band fluxes (whenever possible) of all the RL AGNs already published with a redshift above 4.5.
All these data will be included in a forthcoming work (Belladitta et al. in prep), where we will study the multiwavelength properties of the RL AGN population above z=4.5.

\subsection{Radio properties}
\label{radioprop}
\begin{figure} 
        \centering
        \includegraphics[width=9.0cm,height=9.5cm]{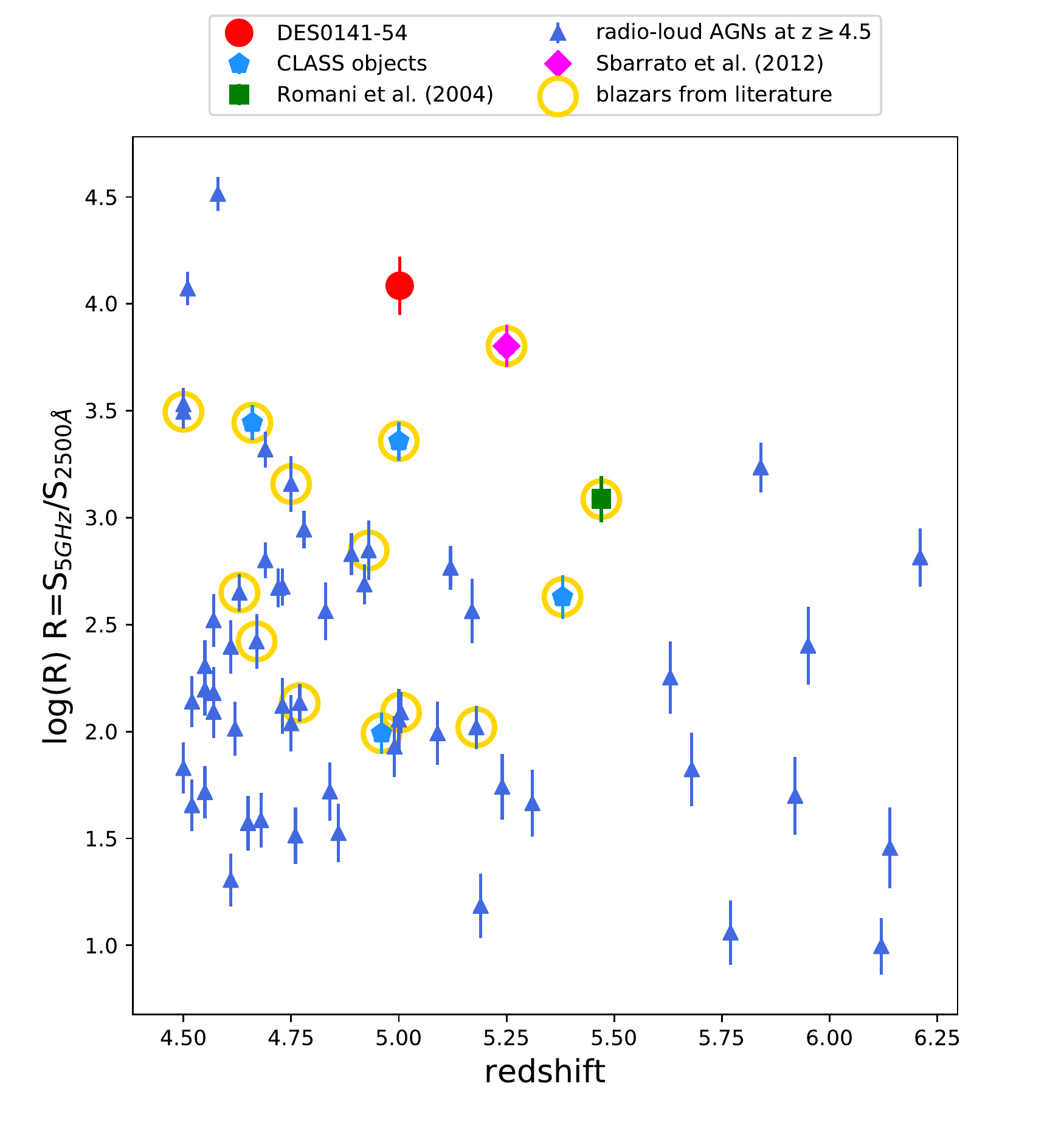}
                \vskip -0.5cm
        \caption{Radio-loudness (R) comparison between DES0141-54 (red point) and high-z RL AGN sample from literature (blue triangles). DES0141-54 has the highest R at z$\geq$5. It is higher than those of the three most distant blazars already discovered (J0906+6930, J1648+4603 and J1026+2542). The objects circled in yellow are all the confirmed blazars found in the literature. We point out that these are only the blazars to have been validated in the literature.}
        \label{radioopt}
\end{figure}

\begin{figure}
        \centering
        \includegraphics[width=9.0cm]{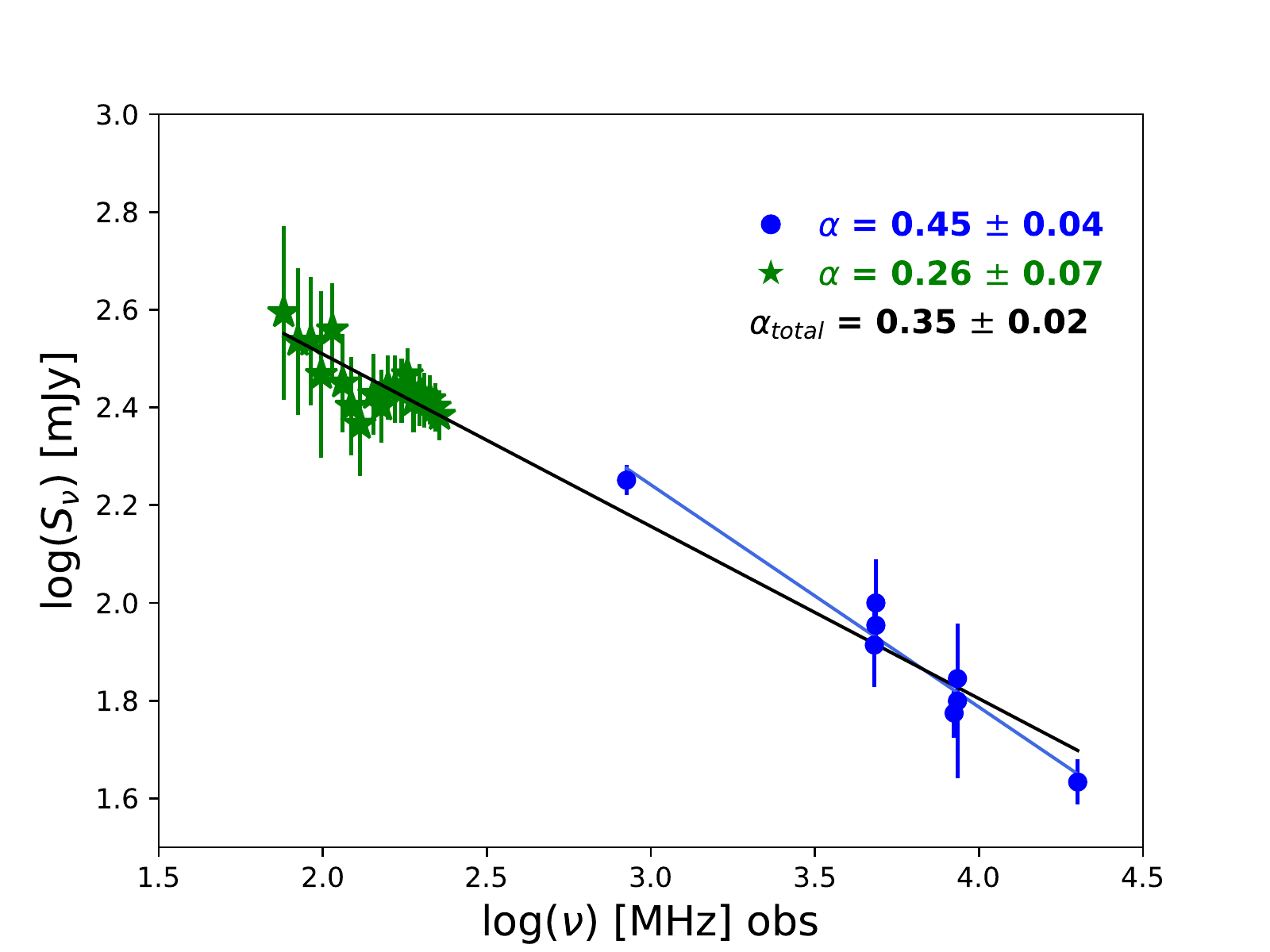}
        \caption{Radio spectrum (observed frame) of DES0141-54. Green stars are the GLEAM low frequency detections, while blue points include radio detections from 0.8 GHz (SUMSS) to 20 GHz (AT20G). The corresponding indexes are reported in the legend. Finally the total radio spectral index ($\alpha_{total}$) computed over the whole range of frequencies is reported in black.} 
        \label{radiospec}
\end{figure}

From the observed optical and radio flux densities we calculated the radio–loudness (R) of DES0141-54 following the definition of Shen et al. (2011): R = S$_{5GHz}$/S$_{2500\AA}$ (rest frame). 
This parameter describes the level of power of the non-thermal synchrotron emission with respect to the thermal one. 
Therefore, a high R supports the idea that the radio emission is boosted along our line of sight.
The rest frame frequency at 5 GHz of DES0141-54 corresponds to an observing frequency of $\sim$0.83 GHz, therefore similar to the SUMSS frequency.
Instead, the rest frame flux density at 2500Å was determined starting from the flux density in the $J$ band and using a power-law continuum (S$_{\lambda} \propto$ $\lambda^{\alpha_{\lambda}}$ ) with the spectral index computed from the DES and WISE photometric points ($\alpha_{\lambda}$=-1.2). 
We obtained a radio-loudness >10$^{4}$ (R=12000$\pm$1600). \\
Figure \ref{radioopt} shows the comparison between the radio-loudness of DES0141-54 and that of high-z RL AGNs from the literature.
For all these objects the radio-loudness was calculated with the definition of Shen et al. (2011) used for DES0141-54\footnote{For these RL AGNs the flux at 5 GHz rest frame was computed starting from the observed flux at 1.4 GHz, found in the reference papers. The radio spectral index was reported in the literature only for few sources, therefore, for the remaining majority we assumed two different spectral indexes, ($\alpha_{\nu}$ = 0.0 and 0.7) and we computed the rest frame flux as the mean of the two. 
Instead the flux at 2500Å rest frame was computed from the $z$ band magnitude (the one available for all these sources) and assuming different spectral indexes ($\alpha_{\lambda}$=-1.7 from Selsing et al. 2016 and $\alpha_{\lambda}$=-1.56 from Vanden Berk et al. 2001).}. 
Figure \ref{radioopt} shows that DES0141-54 has the highest R ever measured at redshift above five. 
From Fig. \ref{radioopt} we also noticed that the majority of blazars found in the literature occupies the portion of the plot at very high R (64\% have log(R)>2.5), suggesting that DES0141-54 has a high probability of being a blazar.
However there are two objects with R greater than or equal to that of DES0141-54: J0311+0507 (Kopylov et al. 2006) at z=4.51 and J2102+6015 (Sowards-Emmerd et al. 2004) at z=4.57 respectively. 
For both of them no evidence of Doppler boosting from Very Long Baseline Interferometry (VLBI) observations is reported in the literature, therefore these sources are not classified as blazars.
The first one is a complex multicomponent ultra steep spectrum radio source (Parijskij et al. 2014).
The second one is classified as a Giga-Hertz Peaked Source (GPS, O'Dea 1998), meaning a very young and compact radio galaxy resolved only at VLBI scale with a peaked radio spectrum above 1 GHz (Frey et al. 2018). 
Therefore, although the high radio-loudness supports the blazar nature of DES0141-54, we cannot exclude the possibility that it is an intrinsically powerful misaligned RL AGN, like J2102+60 and J0311+0507.
For this reason we studied the radio spectrum of DES0141-54 to test for a possible GPS nature.
Thanks to the large number of radio detections, we computed the wide band radio spectrum, from 76 MHz up to 20 GHz, of DES0141-54 (Fig. \ref{radiospec}). 
We fitted all the photometric radio points with a single power-law (S$_{\nu} \propto$ $\nu^{-\alpha_{\nu}}$) to measure the radio spectral slope. 
The radio slope between 0.8 and 20 GHz is $\alpha$=0.45$\pm$0.04. 
By also taking into account the low frequency GaLactic and Extragalactic All-sky MWA Survey (GLEAM) detections, $\alpha$ becomes equal to 0.35$\pm$0.02.
Therefore, the radio spectrum can be considered flat ($\alpha$<0.5).
Moreover, the radio slope reported in the GLEAM catalog for this object, computed using only low frequency data, is $\alpha$=0.259$\pm$0.068. This confirms that the flatness of the spectrum also occurs at low frequencies.
These radio spectral properties make the GPS hypothesis very unlikely and support, instead, the idea that DES0141-54 is a beaming dominated object (i.e., a blazar).\\
This hypothesis is also validated by radio variability of DES0141-54 at 8.6 GHz and at 4.8 GHz, a characteristic typical of blazar sources. 
In fact, as shown in Fig. \ref{radiospec} and Table \ref{Tradio}, DES0141-54 was observed in different years at these two frequencies. 
This flux variation ($\sim$10\% at both frequencies) is consistent with that observed in other high-z blazars ($\sim$14\%, Caccianiga et al. 2019).\\

\subsection{X-ray properties}
\label{xrayprop}
\begin{figure} 
        \centering
        \includegraphics[width=9.5cm,height=9.5cm]{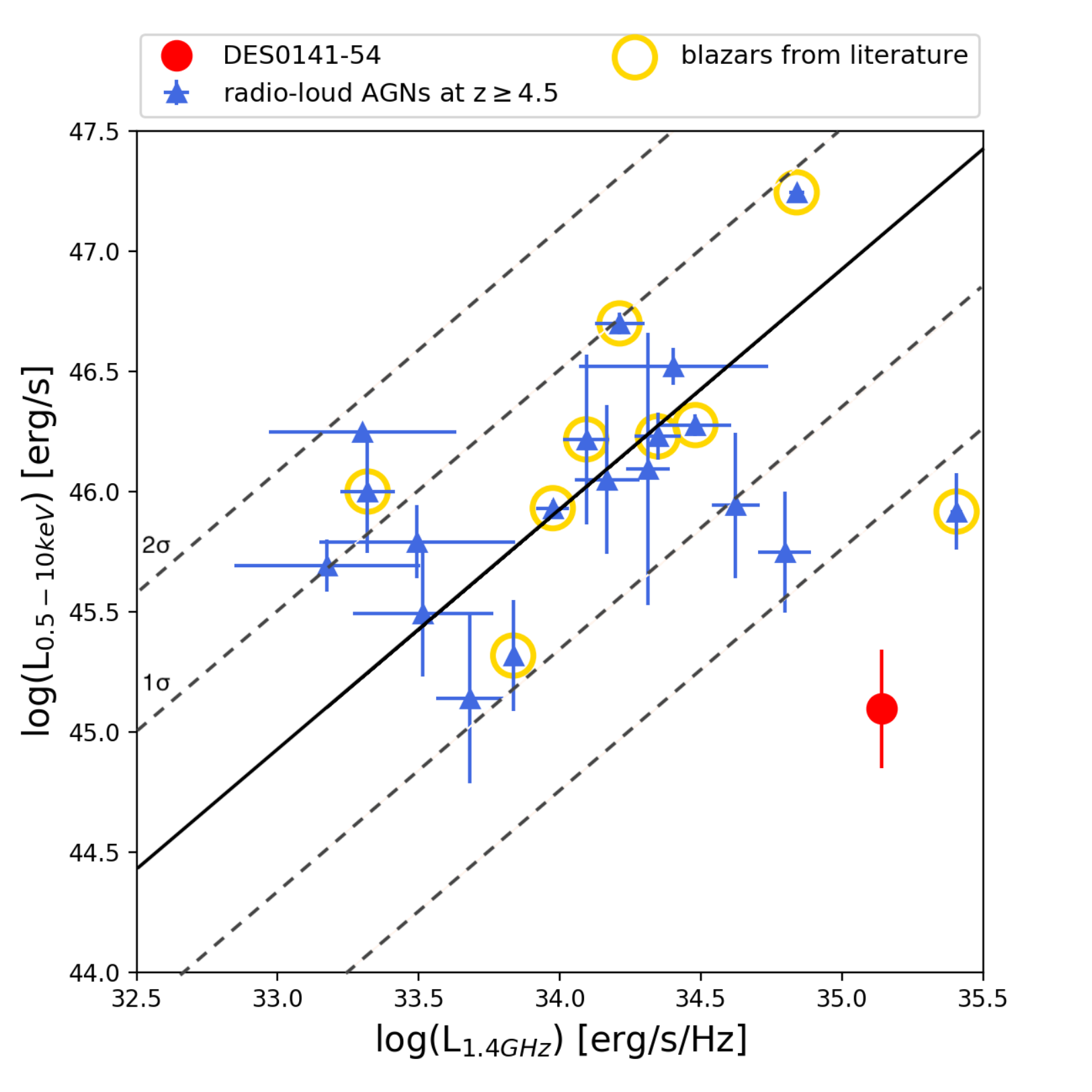}
        \caption{Radio versus X--ray luminosity of DES0141-54 (red point) compared to RL high-z quasars (blue triangles). Yellow circles indicate the confirmed blazars found in the literature, as in Fig. \ref{radioopt}. The solid black line is the 1:1 relation, and the two dashed black lines are the 1$\sigma$ and 2$\sigma$ relations. DES0141-54 is one of the most powerful radio high-z AGNs ever discovered though it has a very weak X--ray luminosity (>2$\sigma$ out of the mean distribution).} 
        \label{XrayComparison}
\end{figure}

\begin{figure} 
        \centering
        \includegraphics[width=9.0cm,height=9.5cm]{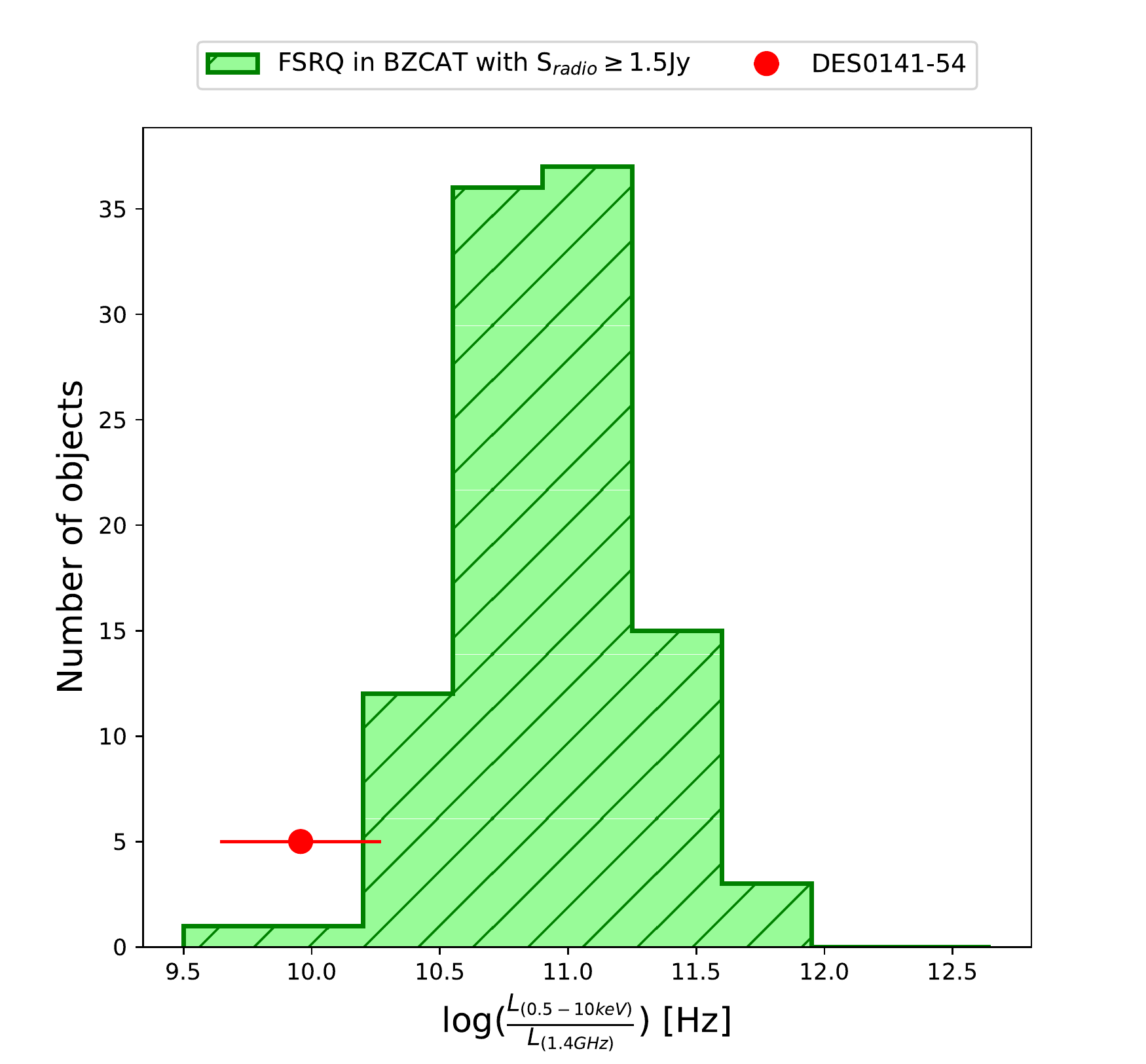}
        \caption{Distribution of XR for FSRQs with radio flux greater than 1.5 Jy in BZCAT. DES0141-54 XR (red point) is similar (within 1$\sigma$) to those of a small number (i.e., a tail) of low-z blazars. These similarities suggests that DES0141-54 can be the high-z analog of such rare low-z blazars.}
        \label{Xradio_histo}
\end{figure}

The X--ray flux computed from the XMM-Newton archival image, and the upper limit obtained from our $Swift$-XRT observation, indicate that DES0141-54 is one order of magnitude fainter in the X--rays compared to other confirmed blazars at z$\geq$5.0 (see, e.g., Romani et al. 2004, Sbarrato et al. 2012 for examples).
To quantify this difference, in Fig. \ref{XrayComparison} we compare the X-ray luminosity at [0.5-10]keV of DES0141-54 with that of the same sample of z>4.5 RL AGNs from the literature used for the radio comparison.
Figure \ref{XrayComparison} shows the X--ray luminosity as a function of 1.4 GHz luminosity for this sample. 
DES0141-54 is in the low X--ray luminosity end of the distribution, with an X--ray luminosity >2$\sigma$ lower than the average of this sample.
The statistics of this sample is low because only a few of these high-z objects have been observed in the X--rays up to now. 
In forthcoming work (Ighina et al. in press) we will study the X--ray and radio properties of a statistical sample of high-z blazars  (the CLASS sample from Caccianiga et al. 2019) and we will establish how common such "X--ray weak" blazars are at high-z.\\
Moreover we note that the X--ray luminosity of DES0141-54 is consistent within 1$\sigma$ with that expected from a radio-quiet quasar with the same optical luminosity (based on the Strateva et al. 2005 relation). 
This suggests that the X-ray emission from the relativistic jet must be very weak.\\
Since DES0141-54 is the only RL AGN above z=4.5 observed up to now that has a very high radio luminosity but a very weak X--ray luminosity, we looked for similar sources at lower redshifts.
To this end, we used the Roma BZCAT-5th edition (Massaro et al. 2015) that collects most of the confirmed blazars discovered so far.
In particular, we considered only the flat spectrum radio quasars (FSRQs) of the BZCAT with a radio flux density (0.843 or 1.4 GHz) above 1.5 Jy. 
This "radio cut" has many advantages: first, at such high radio fluxes the census of blazars is reasonably complete and the sources can be considered a well-defined radio flux-limited sample. 
Second, all these objects have been observed and detected in the X--rays which excludes any possible bias against particularly X--ray weak sources. 
Finally, the selected objects have radio luminosities similar to DES0141-54 (L$_{1.4 GHz}$=1.38$\pm$0.11$\times$10$^{35}$ erg s$^{-1}$ cm$^{-2}$ Hz$^{-1}$), making the comparison more straightforward.
With this selection criteria we obtained a sample of 105 FSRQs with a mean redshift of 1.24.
For all these objects the X-ray flux reported in the BZCAT at [0.1-2.4]keV was converted to [0.5-10]keV assuming an X--ray spectral index ($\alpha_X$) equal to 0.5.
From this flux, we computed the X--ray luminosity in the same energy band. 
For the k-correction in the radio band we assumed $\alpha_r$=0.0. 
In Fig. \ref{Xradio_histo} we show the distribution of the X--ray-to-radio luminosity ratio (XR) for these BZCAT blazars. 
We overplotted the value derived for DES0141-54: log(XR) = 9.96$\pm$0.30 Hz.
This figure shows that blazars with such low XR values are also very uncommon at low-z and correspond to the tail of the distribution.
In particular, only two blazars (i.e., $\sim$2\% of the total sample) have an XR similar to DES0141-54.
The possible origin of this type of X--ray weakness will be discussed in Section \ref{sedmodel}. 

\section{Black hole mass estimates}
\label{mass}
We computed the central black hole mass (M$_{\rm BH}$) of DES0141-54 following two different methods. The first method is the commonly used virial approach and the second method is based on the accretion disk emission.

\subsection{Virial mass}
\label{BH}
We estimated M$_{\rm BH}$ from our single-epoch VIS and NIR X-Shooter spectra by using the broad C$\rm{IV}$$\lambda$1549Å and Mg$\rm{II}$$\lambda$2798Å emission lines.  
Although the C $\rm{IV}$ line is not the best virial estimator (e.g. Richards et al. 2011, Denney 2012), we decided to use it in order to provide a second, independent, M$_{\rm BH}$ estimate in addition to the one based on the very noisy Mg $\rm{II}$ line.\\
We derived M$_{\rm BH}$ following Shen et al. (2011):
\begin{equation}
$log($\frac{M_{\rm BH}}{M_{\odot}}$) = $a$ + $b$log($\frac{\rm \lambda L_{\lambda}}{10^{44}erg/s}$)  + 2log($\frac{\rm FWHM}{\rm km/s}$)$, 
\end{equation}
where the coefficients $a$ and $b$ are empirically calibrated against local AGNs with the reverberation mapping technique (e.g. Peterson et al. 2004). 
In this work we used (a, b) = (0.660, 0.53) for C $\rm{IV}$ (Vestergaard \& Peterson 2006) and (a, b) = (0.740, 0.62) for Mg $\rm{II}$ Shen et al. (2011) respectively.\\
We computed the FWHMs of the C $\rm{IV}$ and Mg $\rm{II}$ emission lines through a single Gaussian spectral fit. 
The rest frame continuum luminosities ($\lambda L_{\lambda}$) at 1350Å and 3000Å have been estimated respectively from the $i$-band and $J$-band flux densities, assuming a single power-law continuum (f$_{\lambda} \propto$ $\lambda^{\alpha_{\lambda}}$) with a slope of $\alpha_{\lambda}$=-1.2, computed from the DES and WISE photometric points. 
The FWHM, the continuum luminosity and the computed M$_{\rm BH}$ values are reported in Table \ref{Tmbh}. 
These M$_{\rm BH}$ virial estimates are affected by large systematic uncertainties ($\geq$0.4 dex, e.g. Shen et al. 2008, 2009).
Moreover, as stressed by Denney et al. (2013), the  Vestergaard \& Peterson (2006) relation based on the FWHM of the C $\rm{IV}$ line, may systematically underestimate the black hole mass.\\
Following this, we derived the Eddington luminosity (L$_{Edd}$) and the Eddington ratio ($\rm \lambda_{Edd}$) based on M$_{\rm BH}$. 
We first estimated L$_{bol}$ using the following bolometric corrections: 
\begin{align}
L_{bol} & = K_{a} \times L_{1350} \notag\\
L_{bol} & = K_{b} \times L_{3000} \\
L_{bol} & = K_{c} \times L_{5100}, \notag 
\end{align}
where L$_{1350}$, L$_{3000}$ and L$_{5100}$ are the continuum luminosities at 1350Å, 3000Å and 5100Å respectively, and $K$ are correction factors. K$_{a}$ is from Shen et al. (2008) and $K_{b}$ and $K_{c}$ are from Richards et al. (2006).  
The estimated values of L$_{bol}$ and their errors, which already take into account the uncertainty on $K$, are reported in Table \ref{Tmbh}.
We then used a mean value of the bolometric luminosity (<L$_{bol}$>=5.2$\pm$1.7$\times$10$^{46}$ erg s$^{-1}$) to calculate the mean $\rm \lambda_{Edd}$ ratio from C $\rm{IV}$ and Mg $\rm{II}$ lines. 
We obtain $\rm \lambda_{Edd\_C \rm{IV}}$ = 1.8$\pm$0.7 and $\rm \lambda_{Edd\_Mg \rm{II}}$ = 0.9$\pm$0.3.
These values indicate that DES0141-54 is accreting very close to the Eddington limit, in agreement with the theoretical models that suggest very high accretion rates (close or even above the Eddington limit) to explain the presence of SMBH at high redshifts (e.g. Lupi et al. 2016, Pezzulli et al. 2016).\\
Finally, from <L$_{bol}$> we computed the luminosity emitted isotropically from the accretion disk, following the relation found in Calderone et al. (2013): L$_{bol}$$\sim$2L$_{disk}$.
We obtain L$_{disk}$$\sim$2.6$\pm$0.8$\times$10$^{46}$ erg s$^{-1}$.
In order to check the consistency of our results, we also estimated the isotropic disk luminosity from the line luminosities of Ly-$\alpha$, N $\rm{V}$, C $\rm{IV}$ and Mg $\rm{II}$ (see Table \ref{Tmbh}), by fitting them with a single Gaussian component. 
Following Celotti et al. (1997) (who followed Francis et al. 1991), we computed the total luminosity of the broadline region (BLR) from the C $\rm{IV}$ and Mg $\rm{II}$ lines, obtaining L$_{BLR}$ = 1.1$\pm$0.7$\times$10$^{45}$ erg s$^{-1}$.
If we assume that the BLR intercepts 10\% of the disk luminosity (hence a covering factor of 0.1), we estimated an L$_{disk}$ equal to  1.1$\pm$0.9$\times$10$^{46}$ erg s$^{-1}$, which is consistent within the errors with our previous estimate based on <L$_{bol}$>.
The error on the disk luminosity does not take into account the uncertainties due to the unknown covering factor.

\begin{footnotesize}
        \begin{table}[!h]
                \caption{Optical and IR properties of DES0141-54.}
                \label{Tmbh}
                \centering
                \begin{tabular}{lc}
                        \hline\hline
                        \multicolumn{2}{c}{Parameters related to C $\rm{IV}$ line} \\
                        \hline
                        FWHM & 1934$\pm$31 \\
                        $\lambda$L$_{\lambda}$ at 1350$\mbox{\AA}$ &  1.2$\pm$0.1 \\
                        M$_{\rm BH}$ &  2.18$\pm$0.13 \\
                        \hline
                        \multicolumn{2}{c}{Parameters related to Mg $\rm{II}$ line} \\
                        \hline
                        FWHM & 2447$\pm$141 \\
                        $\lambda$L$_{\lambda}$ at 3000$\mbox{\AA}$ &  7.3$\pm$1.7 \\
                        M$_{\rm BH}$ & 4.7$\pm$1.6  \\
                        \hline 
                        \multicolumn{2}{c}{Bolometric luminosities (10$^{46}$ erg/s)} \\
                        \hline
                        L$_{bol}$ at 1350$\mbox{\AA}$ & 4.7$\pm$1.6 \\
                        L$_{bol}$ at 3000$\mbox{\AA}$ & 3.8$\pm$1.2 \\          
                        L$_{bol}$ at 5100$\mbox{\AA}$ & 7.1$\pm$1.5 \\          
                        \hline
                        \multicolumn{2}{c}{Line luminosities (10$^{44}$ erg/s)} \\
                        \hline
                        Ly-$\alpha$ & 3.33$\pm$0.10  \\
                        N $\rm{V}$ &  3.71$\pm$0.15  \\
                        C $\rm{IV}$ & 0.66$\pm$0.08  \\
                        Mg $\rm{II}$ & 1.23$\pm$0.19 \\
                        \hline
                        
                \end{tabular}
        
                \tablefoot{The FWHM are reported in km/s; $\lambda$L$_{\lambda}$ is the continuum luminosity near the C $\rm{IV}$ (1350$\mbox{\AA}$) and the Mg $\rm{II}$ (3000$\mbox{\AA}$) lines in unit of 10$^{46}$ erg/s; the black hole mass is in unit of 10$^{8}$ M$_{\odot}$.}
                \end{table}
        \end{footnotesize}

\subsection{Accretion disk models}
\label{sb}
\begin{figure}
\centering
\includegraphics[width=9.0cm]{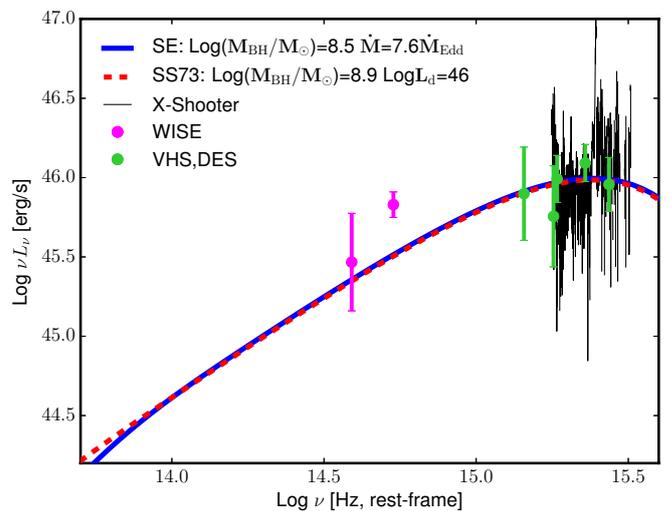}
\caption{Accretion disk models of optical spectrum and optical--IR photometric data of DES0141-54. Dashed red line represents the best modeling of SS73, instead the blue line represents the best model from an SE model. 
We corrected the spectrum and the photometric points for absorption due to intervening H$\rm I$ clouds using the model described in Meiksin (2006).}
\label{mass_fit}
\end{figure}

The optical and IR data can be used to estimate the black-hole mass and the accretion rate independently from virial methods. 
More specifically, we used two models for the accretion disk emission: the standard Shakura \& Sunyaev model (SS73, 1973) and a super-Eddington model (SE, Ohsuga et al. 2002), which is more suited to sources that accrete close to the Eddington limit. 
As a first test, we considered the emission from a standard geometrically thin, optically thick accretion disk (SS73) seen face-on. 
Assuming a non--spinning\footnote{Usually RL AGNs are associated to spinning black holes. However assuming a non-spinning black hole is typical for an SS73 model and is also justified by the results of Campitiello et al. (2018) who found an equivalence between the accretion disk fit with an SS73 model and a KerrBB model with spin$\sim$0.8 and observed face-on (as expected for blazars).} black hole, the disk inner radius corresponds to the innermost stable circular orbit: $R_{\rm in}=3R_{\rm Schw}=6GM_{\rm BH}/c^2$. 
The model that better represents our data is shown in Fig. \ref{mass_fit} and describes the emission around a black hole with M$\rm _{BH}$= $8\times10^8$M$_\odot$ from a disk with a luminosity of L$_{\rm d}=10^{46}$ erg s$^{-1}$.\\
The SS73 model is very well known, but has strong limitations. 
The main problem is related to the possibility for high redshift objects to accrete close to the Eddington limit.  
Ohsuga et al. (2002) introduced a correction to the standard disk geometry when the disk emission is close to $30\%L_{\rm Edd}$.
In this case the radiation pressure affects the geometry and the optical thickness of the disk inner region, hence photons are advected toward the black hole faster than being radiated away. 
This happens within the so--called photon-trapping radius (R$_{\rm pt}$ $\simeq$ $\frac{\dot M}{\dot M_{\rm Edd}}$ R$_{\rm Schw}$, with $\dot M_{\rm Edd}=L_{\rm Edd}/c^2$) where the luminosity emitted by the disk drops dramatically. 
Outside $R_{\rm pt}$, the disk emits as SS73, and therefore the optical--IR SED can be well approximated by a SS73 truncated at $R_{\rm pt}$, if larger than $3R_{\rm Schw}$. 
Assuming this correction to mimic a quasi- or full SE behavior, our data are best described by a black hole of $3\times10^8$M$_\odot$, accreting at $\dot M = 3.5\times10^{26}$ g s$^{-1}$$\simeq8\dot M_{\rm Edd}$ and emitting L$_{\rm d}=10^{46}$ erg s$^{-1}$ $\simeq24\%L_{\rm Edd}$.\\
Figure \ref{mass_fit} shows that the two best models are very similar but correspond to quite different values of M$_{\rm BH}$. 
This difference shows the level of uncertainty on the M$_{\rm BH}$ derived through this method.
However the M$_{\rm BH}$ values computed with these two models are both consistent with the values obtained from the virial method, considering the global uncertainties on the used relations (0.4dex).  
Therefore, we can conclude that the SMBH hosted by DES0141-54 has a mass between 3$\times10^8$ and 8$\times10^8$ M$_{\odot}$.
This makes DES0141-54 the RL AGN that hosts the smallest SMBH discovered up to now at high-z.

\section{Spectral energy distribution (SED) modeling}
\label{sedmodel}
\begin{figure} 
        \centering
        \includegraphics[width=9.0cm, height=11.0cm]{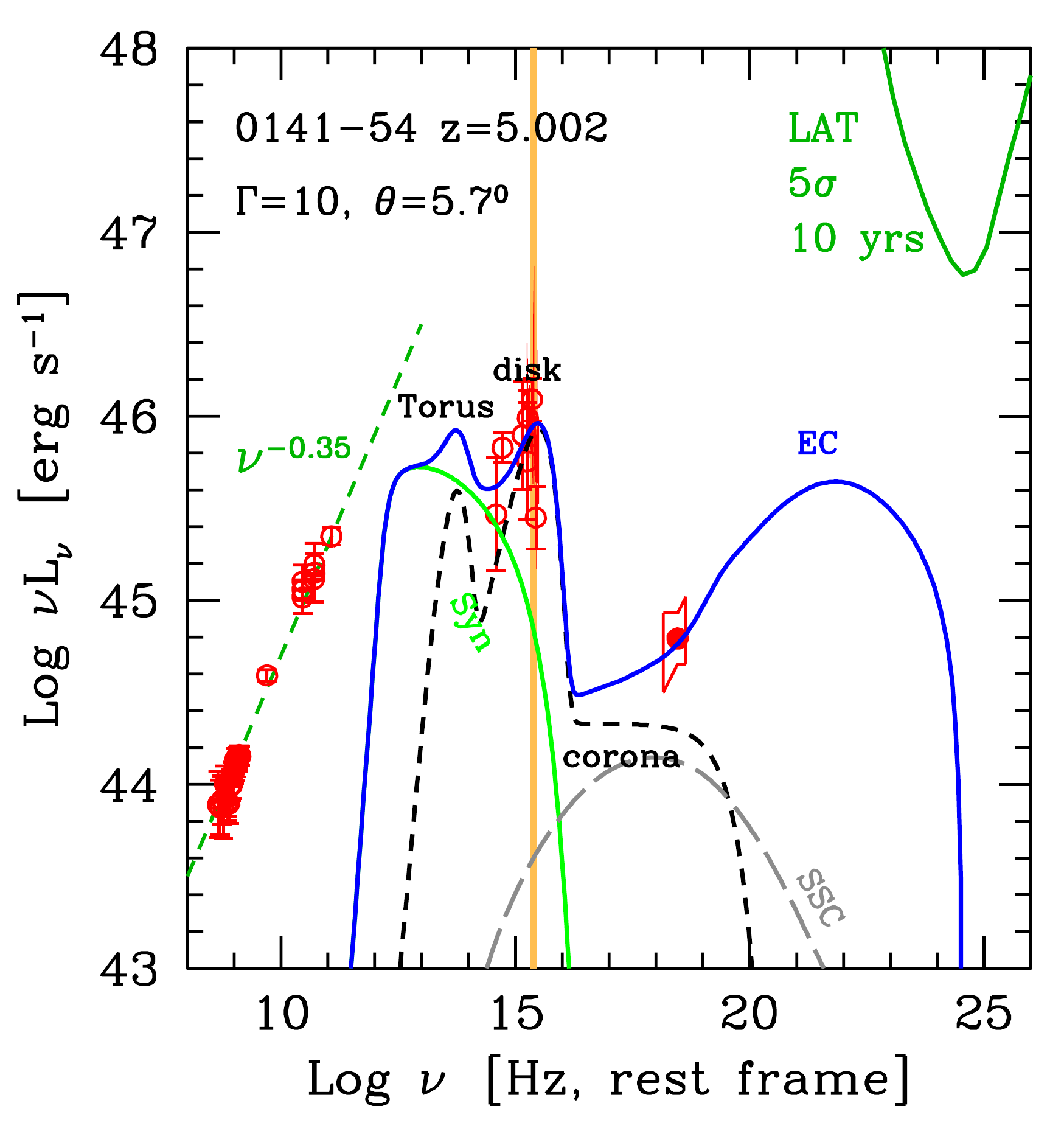} 
        \caption{SED of DES0141-54 from the radio to X--ray frequencies: the solid blue line shows the theoretical blazar model; the dashed black line is the spectrum of an accretion disk, together with the IR torus radiation and the X--ray emission produced in the hot corona; the solid light green line refers to the synchrotron emission; the grey long dashed line is the SSC component. The vertical orange line marks the position of the Ly-$\alpha$ break. As described in Ghisellini \& Tavecchio (2009) the model does not reproduce the radio part of the SED because the jet emitting region that interacts with the external photons is compact and therefore auto--absorbed at low frequencies. We also report in green the $Fermi$LAT sensitivity curve after 10 years (at 5$\sigma$) that shows that our predictions are consistent with a non detection by $Fermi$.}
        \label{sed}
\end{figure}

As a result from our analysis, DES0141-54 is, on the one hand, a very high-z radio powerful FSRQ, which strongly supports its blazar nature; on the other hand, it has an X-ray luminosity that is very low compared to most ($\sim$98\%) of the blazar population, assuming that the comparison BZCAT sample is representative of the entire blazar population in terms of XR. 
Therefore we tried to reproduce the SED of DES0141-54 with the model fully described in Ghisellini \& Tavecchio (2009) to understand whether such a low X--ray luminosity is still compatible with the theoretical predictions of the beaming model.
This is a simple, one-zone, leptonic model, in which relativistic electrons emit through the synchrotron and Inverse Compton processes. 
The electron distribution is derived through a continuity equation, assuming continuous injection, radiative cooling, possible pair production, and pair emission.
Besides the radiation produced internally in the jet, the model also considers radiation coming from the accretion disk, the BLR, the dusty torus and the hot thermal corona sandwiching the accretion disk.\\
The overall rest frame SED\footnote{All the values that can be used to reproduce the SED of DES0141-54 are listed in Table \ref{sedval}.} (in $\nu$L$_{\nu}$ vs $\nu$) of DES0141-54, from radio to X--rays, is shown in Fig. \ref{sed}, together with the model that describes the SED qualitatively, using a small viewing angle ($\theta$), equal to 1/$\Gamma$. 
This is the typical SED of a high-z blazar: the synchrotron emission peaks in the submillimeter--millimeter (sub-mm) band and the Inverse Compton in the GeV band, leaving the accretion disk component ‘`naked'’ and thus observable.
Table \ref{para} lists the parameters of the model.
Some of them are computed independently of the model (M$_{\rm BH}$ from the virial method, L$_{disk}$ from the SS73 disk modeling using the virial mass, R$_{BLR}$ and R$_{torus}$ from L$_{disk}$\footnote{L$_{disk}$: R$_{BLR}$=10$^{17}$$\sqrt{L_{disk,45}}$cm and R$_{torus}$=2.5$\times$10$^{18}$$\sqrt{L_{disk,45}}$cm}) and then used as input parameters. 
The other input parameters (R$_{diss}$, $\Gamma$, $\theta$, $\delta$, $\gamma_b$, $\gamma_{max}$, s$_1$ and s$_2$) are described in detail in Ghisellini \& Tavecchio (2009).
Finally P$_{e}$ and B are the output parameters.
The values used for this model (all the input parameters and P$_{e}$)  are rather typical for FSRQs at high redshift (see Sbarrato et al. 2012, Ghisellini et al. 2015 for examples).
This means that the observed weak X-ray emission is still consistent with the blazar hypothesis.
However to take into account the very powerful radio emission, a high value of the AGN magnetic field is needed. 
In our case B is approximately 9G, which is higher with respect to the typical values (<B> = 4.6G) found in Ghisellini \& Tavecchio (2015) for powerful $Fermi$ blazars.
Indeed only 6\% of these powerful FSRQ have a B equal to or higher than that of DES0141-54.
A large value of the magnetic field is consistent with a relatively small black hole mass, as discussed in Ghisellini \& Tavecchio (2009), which derived that the magnetic energy density U$_B$ $\propto$ 1/M$_{\rm BH}$, and, therefore, B $\propto$ M$^{-1/2}$ (in accordance also with models from Shakura \& Sunyaev 1973 and Novikov \& Thorne 1973). 
This can explain why, in our source, the synchrotron process dominates the overall electromagnetic output. As a consequence, DES0141-54 is expected to be faint in $\gamma$--rays, and not detectable by $Fermi$ as shown in Fig. \ref{sed}.

\begin{table*} 
        \caption{Adopted parameters for jet model.} 
        \label{para}
        \tiny
        \centering
        \begin{tabular}{lllll lllll lllll}
        \hline
                \hline
                M$_{\rm BH}$ &L$_{\rm disk}$ &L$_{\rm disk}$/L$_{\rm Edd}$ &R$_{\rm diss}$ &R$_{\rm BLR}$ &R$_{\rm torus}$ 
                &$P^\prime_{\rm e, 45}$  &B &$\Gamma$ &$\theta_{\rm V}$ &$\delta$ &$\gamma_{\rm b}$ 
                &$\gamma_{\rm max}$ &s$_1$  &s$_2$ \\ 
                ~(1) & (2) & (3) & (4) & (5) & (6) & (7) & (8) & (9) & (10) & (11) & (12) & (13) & (14) & (15) \\
                \hline   
         4$\times$10$^8$  &8.32 &0.16 &120  &288  &5.4e3  &0.13 &9.63  &10 &5.7  &10.07  &2.e2  &3e3 &1 &2.5  \\
                \hline
        \hline  
                \end{tabular}
        \vskip 0.4 true cm
        \tablefoot{Col. (1): black hole mass in solar masses;
                Col. (2): disk luminosity in units of $10^{45}$ erg s$^{-1}$;
                Col. (3): disk luminosity in units of the Eddington luminosity;
                Col. (4): distance of the dissipation region from the black hole, in units of $10^{15}$ cm;
                Col. (5): size of the BLR, in units of $10^{15}$ cm;
                Col. (6): size of the torus, in units of $10^{15}$ cm;
                Col. (7): power injected in the jet in relativistic electrons, calculated in the comoving frame, in units of $10^{45}$ erg s$^{-1}$; 
                Col. (8): magnetic field in G;  
                Col. (9): bulk Lorentz factor;
                Col. (10): viewing angle in degrees;
                Col. (11): relativistic Doppler factor;
                Col. (12): break random Lorentz factor of the injected electrons;
                Col. (13): maximum random Lorentz factor of the injected electrons;
                Col. (14): slope of the injected electron distribution before the break;
                Col. (15): slope of the injected electron distribution beyond the break.
        }
\end{table*}

\section{Summary and conclusions}
\label{sumconc}
In this work we presented the discovery and the multiwavelength properties of DES0141-54, a very powerful RL AGN at high redshift (z=5.0) when the Universe was only 1.18 billion years old. 
This source was discovered by cross-matching the first data release of the DES with the SUMSS radio catalog. 
This is the first RL quasar discovered in the DES at high redshift.
DES0141-54 is characterized by a flat radio spectrum and a high radio-loudness, which are typical features of blazars, but it has a weak X--ray emission compared to other blazars at low and high redshift.
Therefore DES0141-54 appears to belong to a minority (2\%) of the total blazar population that is characterized by a very weak X-ray luminosity. 
Using a simple beaming model, we have shown that the very powerful radio luminosity and the weak X-ray emission can be explained with a very high value of the magnetic field (B$\sim$9G). 
X--ray data of better quality and at larger energies (above 10 keV observed frame), as well as a long-time scale X--ray monitoring, will be necessary in order to firmly constrain the source orientation and the physical properties of the jet emission.
With DES0141-54 we have increased the number of z$\geq$5.0 blazars known (seven in total, including DES0141-54). \\
Finally, from the analysis of the optical and IR spectrum, and from the disk modeling, we found that DES0141-54 hosts the smallest SMBH (M$_{\rm BH}$ = 3-8$\times$10$^{8}$ M$_{\rm \odot}$) ever discovered up to now at distant redshift for a RL source. 
This confirms that using very deep photometric catalog (like the DES) we can discover less massive, and therefore less luminous, high-z sources.

\begin{acknowledgements}
We thank the anonymous referee for the useful comments.
This work is based on observations made with ESO Telescopes at the La Silla Paranal Observatory under program 0100.A-0606(A) (New Technology Telescope) and program 2100.A-5039 (Very Large Telescope).
We are grateful to the VLT staff for providing DDT observations for this object.
This work also used data from observations with the Neil Gehrels Swift Observatory (ToO request, target ID:10586).
SB, AM and AC acknowledge support from ASI under contracts ASI-INAF n. I/037/12/0 and n.2017-14-H.0 and from INAF under PRIN SKA/CTA FORECaST. 
CC acknowledges funding from the European Union’s Horizon 2020 research and innovation programme under the Marie Sklodowska-Curie grant agreement No 664931.
We thanks prof. D. Dallacasa (University of Bologna, Italy) for helpful comments and discussion, prof. R.W. Romani (Stanford University) for the precious information and Paola Severgnini (INAF-OA Brera) for the interesting comments.\\
This project used public archival data from the Dark Energy Survey (DES). Funding for the DES Projects has been provided by the U.S. Department of Energy, the U.S. National Science Foundation, the Ministry of Science and Education of Spain, the Science and Technology Facilities Council of the United Kingdom, the Higher Education Funding Council for England, the National Center for Supercomputing Applications at the University of Illinois at Urbana–Champaign, the Kavli Institute of Cosmological Physics at the University of Chicago, the Center for Cosmology and Astro-Particle Physics at the Ohio State University, the Mitchell Institute for Fundamental Physics and Astronomy at Texas A\&M University, Financiadora de Estudos e Projetos, Fundação Carlos Chagas Filho de Amparo à Pesquisa do Estado do Rio de Janeiro, Conselho Nacional de Desenvolvimento Científico e Tecnológico and the Ministério da Ciência, Tecnologia e Inovação, the Deutsche Forschungsgemeinschaft and the Collaborating Institutions in the Dark Energy Survey. The Collaborating Institutions are Argonne National Laboratory, the University of California at Santa Cruz, the University of Cambridge, Centro de Investigaciones Enérgeticas, Medioambientales y Tecnológicas–Madrid, the University of Chicago, University College London, the DES-Brazil Consortium, the University of Edinburgh, the Eidgenössische Technische Hochschule (ETH) Zürich, Fermi National Accelerator Laboratory, the University of Illinois at Urbana-Champaign, the Institut de Ciències de l’Espai (IEEC/CSIC), the Institut de Física d’Altes Energies, Lawrence Berkeley National Laboratory, the Ludwig-Maximilians Universität München and the associated Excellence Cluster Universe, the University of Michigan, the National Optical Astronomy Observatory, the University of Nottingham, The Ohio State University, the OzDES Membership Consortium, the University of Pennsylvania, the University of Portsmouth, SLAC National Accelerator Laboratory, Stanford University, the University of Sussex, and Texas A\&M University.\\
This publication makes use of data products from the Wide–field Infrared Survey Explorer, which is a joint project of the University of California, Los Angeles, and the Jet Propulsion Laboratory/Caltech, funded by the National Aeronautics and Space Administration.
This paper includes archived data obtained through the Australia Telescope Online Archive (http://atoa.atnf.csiro.au).
The Australia Telescope Compact Array and the Parkes radio telescope are part of the Australia Telescope National Facility which is funded by the Australian Government for operation as a National Facility managed by CSIRO.
This research has made use of data obtained from the 3XMM XMM-Newton serendipitous source catalog compiled by the 10 institutes of the XMM-Newton Survey Science Centre selected by ESA.
This publication has made use of data and/or software provided by the High Energy Astrophysics Science Archive Research Center (HEASARC), which is a service of the Astrophysics Science Division at NASA/GSFC and the High Energy Astrophysics Division of the Smithsonian Astrophysical Observatory.
IRAF is distributed by the National Optical Astronomy Observatory, which is operated by the Association of Universities for Research in Astronomy (AURA) under a cooperative agreement with the National Science Foundation.
The X--Shooter pipeline is based on the the X--Shooter Data Reduction Library (DRL) developed by Paolo Goldoni, Frederic Royer, Regis Haigron, Laurent Guglielmi, Patrick Francois, Matthew Horrobin and Hector Flores.
This research made use of Astropy, a community-developed core Python package for Astronomy (Astropy Collaboration,2013; http://www.astropy.org).
\end{acknowledgements}

\appendix
\section{Tables description}
Here we report the full description of the archival radio fluxes reported in the GLEAM catalog and the values useful to build the spectral energy distribution of DES0141-54. 

\begin{footnotesize}
        \begin{table}[!h]
                \caption{Radio detections of DES0141-54 found in GLEAM catalog. The observations have been carried out with the Murchison Widefield Array (MWA) telescope in the years 2013 and 2014.}
                \label{Tradio2}
                \centering
                \begin{tabular}{cc}
                        \hline\hline
                        $\nu_{obs}$  & S$_{\nu}$  \\ 
                        (GHz) & (mJy)   \\  
                        (1) & (2)   \\ 
                        \hline 
                        0.231    & 241.9$\pm$12.3   \\               
                        
                        0.220         &  251.4$\pm$12.6  \\               
                        
                        0.212         &  260.5$\pm$13.1  \\               
                        
                        0.204         &  259.9$\pm$14.5   \\               
                        
                        0.197         &  266.2$\pm$16.9  \\               
                        
                        0.189         &  258.3$\pm$16.5   \\               
                        
                        0.181         &  292.7$\pm$16.1  \\               
                        
                        0.174         &  271.8$\pm$18.0   \\               
                        
                        0.166         &  273.9$\pm$18.9   \\               
                        
                        0.158         &  275.8$\pm$18.1   \\               
                        
                        0.151         &  252.9$\pm$18.9  \\ 
                        
                        0.143         &  267.1$\pm$22.0   \\               
                        
                        0.130         &  231.9$\pm$24.4   \\               
                        
                        0.122         &  252.6$\pm$25.3  \\               
                        
                        0.115         &  281.8$\pm$28.3   \\               
                        
                        0.107         &  361.9$\pm$34.8  \\       
                        
                        0.099         &  293.4$\pm$49.9   \\
                        
                        0.092         &  343.3$\pm$44.9   \\               
                         
                        0.084          &  342.9$\pm$51.3   \\               
                        
                        0.076          &  392.5$\pm$69.8   \\               
                        \hline
                        
                \end{tabular}
                \tablefoot{Col(1): observed frequency in GHz; Col(2): integrated flux density in mJy} 
        \end{table}
\end{footnotesize}

\begin{footnotesize}
        \begin{table}[!h]
                \caption{Frequencies and luminosity for computing the spectral energy distribution of DES0141-54 from radio to X--ray} 
                \label{sedval}
                \centering
                \begin{tabular}{ccc}
                        \hline\hline
    log($\nu_{rest}$)   &  log($\nu$ L$_{\nu}$)  &  survey \\
    (Hz)                &   (erg/s)              &         \\
    (1)                 &   (2)                  &  (3)    \\
    \hline
8.66            &  43.89$\pm$0.18        &  GLEAM \\
8.70            &  43.87$\pm$0.15        &  GLEAM \\
8.74            &  43.91$\pm$0.13        &  GLEAM \\
8.77            &  43.88$\pm$0.17        &  GLEAM \\
8.81            &  44.00$\pm$0.10        &  GLEAM \\
8.84            &  43.93$\pm$0.10        &  GLEAM \\
8.86            &  43.90$\pm$0.10        &  GLEAM \\
8.89            &  43.89$\pm$0.10        &  GLEAM \\
8.93            &  43.99$\pm$0.08        &  GLEAM \\
8.96            &  43.99$\pm$0.07        &  GLEAM \\
8.98            &  44.05$\pm$0.06        &  GLEAM \\
9.00            &  44.07$\pm$0.07        &  GLEAM \\
9.02            &  44.09$\pm$0.07        &  GLEAM \\
9.03            &  44.14$\pm$0.05        &  GLEAM \\
9.05            &  44.10$\pm$0.06        &  GLEAM \\
9.07            &  44.13$\pm$0.06        &  GLEAM \\
9.08            &  44.14$\pm$0.03        &  GLEAM \\
9.09            &  44.14$\pm$0.05        &  GLEAM \\
9.10            &  44.16$\pm$0.05        &  GLEAM \\
9.12            &  44.16$\pm$0.05        &  GLEAM \\
9.13            &  44.15$\pm$0.05        &  GLEAM \\
9.70            &  44.59$\pm$0.03        &  SUMSS \\
10.46           &  45.01$\pm$0.08        &  ATPMN \\
10.46           &  45.05$\pm$0.05        &  AT20G  \\
10.46           &  45.10$\pm$0.09        &  PMN   \\
10.70           &  45.11$\pm$0.05        &  CRATES \\
10.71           &  45.15$\pm$0.16        &  ATPMN \\
10.71           &  45.19$\pm$0.06        &  AT20G \\
11.08           &  45.35$\pm$0.05        &  AT20G \\
14.59           &  45.47$\pm$0.31        &  WISE \\
14.73           &  45.83$\pm$0.08        &  WISE \\
15.16           &  45.90$\pm$0.29        &  VHS  \\
15.25           &  45.75$\pm$0.32        &  DES  \\
15.26           &  45.99$\pm$0.15        &  DES  \\
15.36           &  46.09$\pm$0.12        &  DES  \\
15.44           &  45.45$\pm$0.17        &  DES  \\
18.46           &  44.79$\pm$0.15        &  XMM  \\
                \hline
        \end{tabular}
\tablefoot{Col (1): 
        logarithm of the rest-frame frequency in Hz; col(2): 
        logarithm of the rest-frame $\nu$L$_{\nu}$ in erg/s with their errors; col(3): reference survey.
We report the XMM-Newton point at 2 keV rest-frame.}
\end{table}
\end{footnotesize}

\end{document}